\documentclass[prd,onecolumn,aps,showpacs,nofootinbib,nobibnotes,superscriptaddress,preprintnumbers]{revtex4}
\usepackage{graphicx}
\usepackage{bm}
\usepackage{amsmath}
\usepackage{amssymb}
\usepackage{amsthm}
\usepackage{color}
\usepackage{hyperref}
 \usepackage[utf8]{inputenc}

\newcommand{\diff}{{\rm d}}
\def\d{\mathrm{d}}

\newcommand{\mR}{\mathcal{R}}
\newcommand{\mQ}{\mathcal{Q}}
\newcommand{\mP}{\mathcal{P}}
\newcommand{\divA}{\nabla \cdot A}
\newcommand{\hnabla}{\hat{ \nabla}}
\newcommand{\hGam}{\hat{ \Gamma}}

\newcommand{\dA}{\delta A}
\newcommand{\Fd}{\tilde{F}}

\newcommand{\al}{\alpha}
\newcommand{\bt}{\beta}
\newcommand{\ga}{\gamma}

\newcommand{\la}{\lambda}

\newcommand{\be}{\begin{equation}}
\newcommand{\ee}{\end{equation}}

\newcommand{\lp}{\left(}
\newcommand{\rp}{\right)}

\newcommand{\mS}{{\mathcal S}}

\def\mpl{M_{\rm Pl}}

\newcommand{\bea}{\begin{eqnarray}}
\newcommand{\eea}{\end{eqnarray}}

\newcommand{\lsim}   {\mathrel{\mathop{\kern 0pt \rlap
  {\raise.2ex\hbox{$<$}}}
  \lower.9ex\hbox{\kern-.190em $\sim$}}}
\newcommand{\gsim}   {\mathrel{\mathop{\kern 0pt \rlap
  {\raise.2ex\hbox{$>$}}}
  \lower.9ex\hbox{\kern-.190em $\sim$}}}

\begin{document}
%\hspace{5.2in} \mbox{NORDITA-}\\\vspace{1.53cm} 

\title{Cosmology for quadratic gravity in generalized Weyl geometry}

\date{\today,~ $ $}

\author{Jose Beltr\'an Jim\'enez}
\email{jose.beltran@cpt.univ-mrs.fr}
\affiliation{CPT, Aix Marseille Universit\'e, UMR 7332, 13288 Marseille,  France.}

\author{Lavinia Heisenberg}
\email{lavinia.heisenberg@eth-its.ethz.ch}
\affiliation{Institute for Theoretical Studies, ETH Zurich, Clausiusstrasse 47, 8092 Zurich, Switzerland}

\author{Tomi S. Koivisto}
\email{tomi.koivisto@nordita.org}
\affiliation{Nordita, KTH Royal Institute of Technology and Stockholm University, Roslagstullsbacken 23, 10691 Stockholm, Sweden}

\date{\today}

\begin{abstract}

A class of vector-tensor theories arises naturally in the framework of quadratic gravity in spacetimes with linear vector distortion. 
Requiring the absence of ghosts for the vector field imposes an interesting condition on the allowed connections with vector distortion:
the resulting one-parameter family of connections generalises the usual Weyl geometry with polar torsion. The cosmology of this
class of theories is studied, focusing on isotropic solutions wherein the vector field is dominated by the temporal component. De Sitter
attractors are found and inhomogeneous perturbations around such backgrounds are analysed. In particular, further constraints on the
models are imposed by excluding pathologies in the scalar, vector and tensor fluctuations. Various exact background 
solutions are presented, describing a constant and an evolving dark energy, a bounce and a self-tuning de Sitter phase. 
However, the latter two scenarios are not viable under a closer scrutiny.

\end{abstract}
\preprint{NORDITA-2016-10}

\maketitle

\section{Introduction}

General Relativity (GR) is the most widely accepted theory of gravity. It has been subjected to intense experimental confrontation, but yet there is no
direct evidence that would signal a need for modifications of the theory \cite{Will:2014kxa}. Nevertheless, research of different gravity theories is flourishing. This is partially because GR is not found to be theoretically satisfactory: it is not a quantum theory, and it predicts singularities, for example. Also, coupled with the Standard Model of elementary particles, it predicts a catastrophically large cosmological constant, if one does not resort to technically highly unnatural fine-tuning that is unstable against quantum corrections \cite{Weinberg:1988cp,Martin:2012bt}. Furthermore, if we accept that GR is valid at all scales, we then need to invoke new ingredients in the cosmological matter sector: the notorious dark energy and the dark matter that has escaped all our attempts at direct or indirect detection \cite{Bull:2015stt,Durrer:2007re,Blanchet:2015sra,Blanchet:2015bia}.
 \newline

A defining property of GR is the unity of the metric and the affine structures of spacetime. In GR, the latter is tied to be fully determined as the Christoffel symbols of the metric tensor which is thus the only independent field. Most modifications of GR have been introduced in this context, by considering then more general actions for the metric and by introducing additional fields that couple non-minimally to the metric. The cosmological constant problem, for example, might be addressed by devising an action for the metric that would degravitate ~\cite{Dvali:2002fz, ArkaniHamed:2002fu,Dvali:2007kt, deRham:2010tw}, adjust suitably the underlying symmetries such as by introducing unimodularity  \cite{PhysRevD.40.1048,Shaposhnikov:2008xb} or feature self-tuning with or without extra fields \cite{Kehagias:2000dg,Charmousis:2011bf,deRham:2011by}. On the other hand, the ultraviolet singularities and the obstacles to renormalisation could be smoothened by quantum effects or by higher-derivative terms for the metric in the gravitational action \cite{Stelle:1976gc,Bonanno:2000ep,Biswas:2011ar,Bambi:2013caa}.

In the more general context of non-Riemannian geometries, richer possilibilities emerge as the connection may carry also torsion \cite{Hammond:2002rm} and non-metricity \cite{Krasnov:2007uu,Sobreiro:2007cm,Vitagliano:2013rna}. As also Einstein himself argued, the connection appears to have a more fundamental ontological status than the metric field \cite{einstein2001meaning}, which has become apparent especially in the developments of gauge theories of gravity \cite{Hehl:1976kj,Blagojevic:2002du}. In the Palatini approach, wherein one regards the spacetime connection as an independent field besides the metric, it is well known that GR is dynamically recovered for the Einstein-Hilbert action (and all the subsequent Lovelock invariants \cite{Exirifard:2007da,Borunda:2008kf}), but more general actions become different theories when subjected either to the metric or to the Palatini variation \cite{Olmo:2011uz,Amendola:2010bk,Capozziello:2015lza,Jimenez:2014fla}. A perhaps more conservative than the Palatini approach is however to consider that only some of the 
degrees of freedom residing in the non-metricity or the torsion sector are physically relevant. A prototype example is the Weyl geometry, wherein the only additional field is the trace of the non-metricity, the so called Weyl vector.

Since the first conception of the idea of gauge symmetry, Weyl geometry has provided an important framework for the development of fundamental theories \cite{Scholz:2011za}. In Weyl's unified theory of electromagnetism and gravity, the principle of relativity applied not only to the choice of reference frames, but also to the choice of local units of length. The corresponding spacetime geometry can accommodate scale-invariance, meaning that physics becomes insensitive to for example the difference of masses of various particles. Though the original formulations of Weyl's unified theory can be abandoned as not viable, the idea that physics at some fundamental level is scale invariant remains alive and appealing. Obviously, such a symmetry should be somehow broken at our low energy world, but it still could provide a key to a solution of the most fundamental problems of gravitational physics, if it is accepted that the elimination of the physical propagation of the conformal mode of GR could redeem gravity from both the ultraviolet singularities and the infrared vacuum's weight \cite{Shaposhnikov:2008xi,'tHooft:2009ms}. It was recently argued that local conformal invariance has to be an exact symmetry and further, broken in a spontaneous manner \cite{doi:10.1142/S0218271815430014}.

In this paper we study cosmologies in an extended Weyl geometry. It is not our aim to formulate a scale-invariant theory (for such, see e.g. \cite{Mannheim:2011ds,Varieschi:2008fc}) but simply explore the cosmological implications of a fundamental gravitational vector mode. It is perhaps surprising that the leading order, quadratic curvature corrections to the gravitational action were elucidated in the context of Weyl geometry only very recently \cite{Jimenez:2014rna,Haghani:2014zra}. The result is interesting as it introduces a novel four-parameter class of potentially viable (at least, ghost-free) vector-tensor theories. We considered an extension of the Weyl geometry wherein the distortion (i.e. the non-Levi-Civita part of the connection) is linear in a vector field in the most general non-derivative way, featuring then two extra terms besides the pure Weyl-type non-metricity \cite{Jimenez:2015fva}.
As one of the special cases, this geometry includes the one-parameter non-metric spaces of Ref. \cite{0264-9381-8-9-004}.  Here we will systematically derive the quadratic curvature theory in the geometry with linear vector distortion (for studies of other quadratic non-metric theories, see e.g. \cite{Heinicke:2005bp,Baekler:2006vw}), and apply the results to study various new cosmological scenarios. Vector field cosmologies have been already studied extensively in the literature, with interest in e.g. dark energy \cite{ArmendarizPicon:2004pm,Koivisto:2008xf,Jimenez:2008au}, inflation \cite{Golovnev:2009ks,Golovnev:2011yc,Solomon:2013iza}, anisotropies \cite{Koivisto:2007bp,Thorsrud:2012mu,Akarsu:2013dva,Koivisto:2014gia}, dual fields \cite{Koivisto:2009ew,Koivisto:2011rm}, Gauss-Bonnet couplings \cite{Oliveros:2015jca,Geng:2015kvs}, screening  \cite{BeltranJimenez:2013fca,DeFelice:2016cri}, a cosmological constant \cite{Jimenez:2008nm,Jimenez:2009dt,Jimenez:2009sv} and stability \cite{EspositoFarese:2009aj,Fleury:2014qfa}. Our framework however suggests a fundamental geometric origin for the possible existence of a cosmic vector, furthermore predicting a rather well-specified vector-tensor action.

We begin in the next Section \ref{geometry} by deriving this action from the most general quadratic-in-curvature theory, on which we then impose restrictions by various consistency requirements (we also take a brief look at possible higher order curvature invariants). Remarkably, it turns out that we are then restricted to the class of generalised Weyl geometry, whose special status was already recognised from another aspect \cite{Jimenez:2015fva}. In Section \ref{ds} we then study the existence of de Sitter solutions, and find out that there are 0-4 such solutions depending on the theory parameters. We go further and analyse the propagation of scalar, vector and tensor fluctuations for generic models in de Sitter background. In Section \ref{cosmology} we then study in more detail some specific analytical solutions that are found for some given parameter combinations. We present both interesting late time (dark energy, cosmological constant) and early time (self-tuned de Sitter with first order phase transition, bouncing solution) scenarios. Finally, we conclude in Section \ref{conclusions} and complete some derivations with details in the Appendix.

\section{Generalizing Weyl geometry: Spacetimes with linear vector distortion}
\label{geometry}

In this section we will start by briefly reviewing the basic properties of Weyl geometry and how it can be generalized to include the most general connection linearly determined by a vector field as introduced in \cite{Jimenez:2015fva}. Then, we will proceed to the construction of gravitational actions based on these geometries.

\subsection{Geometrical framework}
The defining property of Weyl geometry is the breaking of the metricity condition by introducing a vector field as follows
\be
\hnabla_\alpha g_{\mu\nu}=-2A_\alpha g_{\mu\nu}.
\label{nonmetricity}
\ee
This relation is preserved by the transformation  $g_{\mu\nu}\rightarrow e^{2\Lambda(x)}g_{\mu\nu}$ when simultaneously we transform $A_\mu\rightarrow A_\mu-\partial_\mu\Lambda(x)$ and, thus, Weyl geometry is a natural arena to formulate conformally invariant theories, although we will not pursue this here. The above expression can be easily solved for the connection (assuming vanishing torsion) to obtain
\be
\hGam^\alpha_{\beta\gamma}=\frac12g^{\alpha\lambda}\Big(g_{\lambda\gamma,\beta}+g_{\beta\lambda,\gamma}-g_{\beta\gamma,\lambda}\Big)-\left(A^\alpha g_{\beta\gamma}-2A_{(\beta}\delta^\alpha_{\gamma)}\right)\,.
\label{Wconnection}
\ee
We see that in Weyl geometry, the connection acquires a distortion linearly depending on $A_\mu$.

The class of geometries introduced in \cite{Jimenez:2015fva} extends the Weyl geometry by allowing for the most general connection with a distortion tensor linearly determined by a vector field. The form of the desired connection is thus
\be \label{vgamma}
\hat{\Gamma}^\al_{\bt\ga} = \Gamma^\al_{\bt\ga}  
-b_1A^\alpha g_{\beta\gamma}+b_2\delta^\alpha_{(\beta} A_{\gamma)}+b_3\delta^\alpha_{[\beta} A_{\gamma]}\,,
\ee
where $\Gamma^\al_{\bt\ga}$ are the usual Christoffel symbols of the metric and $b_i$ are arbitrary coefficients. This connection gives rise to non-metricity, but also contains the trace vector part of the torsion. Interestingly, although the conformal invariance of the metric compatibility condition is lost by the general connection (\ref{vgamma}),  the extra torsion component allows to recuperate it for more general connections. To see it explicitly, we can compute the covariant derivative of the metric, which is given by
\be
\hat{\nabla}_\mu g_{\alpha\beta}=(b_3-b_2)A_\mu g_{\alpha\beta} + (2b_1-b_2-b_3) A_{(\alpha}g_{\beta)\mu}\,.
\label{nonmetricity2}
\ee
Thus, we recover the Weyl condition (\ref{nonmetricity}) provided we impose $2b_1-b_2-b_3=0$. In the absence of torsion $b_3=0$, we exactly recover the Weyl condition, but the torsion $b_3$ term allows to maintain the gauge invariance of (\ref{nonmetricity2}) for more general connections. It is not difficult to see that for $b_3=2b_1-b_2$, the equation (\ref{nonmetricity2}) remains invariant under $g_{\mu\nu}\rightarrow e^{2\Lambda(x)}g_{\mu\nu}$ while $A_\mu\rightarrow A_\mu+\frac{1}{b_1-b_2}\partial_\mu \Lambda(x)$.
We can now introduce the Riemann tensor of our connection as usual
\be
\mR_{\mu\nu\rho}{}^\alpha\equiv\partial_\nu\hGam^\alpha_{\mu\rho}-\partial_\mu\hGam^\alpha_{\nu\rho}+\hGam^{\alpha}_{\nu\lambda}\hGam^{\lambda}_{\mu\rho}-\hGam^{\alpha}_{\mu\lambda}\hGam^{\nu}_{\nu\rho}\,.
\ee
It is important to keep in mind that, since we have torsion and non-metricity, this Riemann tensor does not have the usual symmetries of Levi-Civita connections, but only the antisymmetry in the first two indices inherited from its definition as a commutator: $\mR_{\mu\nu\rho}{}^\alpha=-\mR_{\nu\mu\rho}{}^\alpha$. In particular, this means that we can construct 3 independent traces, namely: the usual Ricci tensor $\mR_{\mu\nu}\equiv\mR_{\mu\alpha\nu}{}^\alpha$, the co-Ricci tensor $\mP_\mu{}^\alpha\equiv g^{\nu\rho}\mR_{\mu\nu\rho}{}^\alpha$ and the homothetic tensor\footnote{Sometimes referred to as the segmental curvature tensor.} $\mQ_{\mu\nu}\equiv\mR_{\mu\nu\alpha}{}^\alpha$. For the connection given in (\ref{vgamma}), these tensors can be expressed as
\bea
\mR_{\mu\nu}&=&R_{\mu\nu}+\frac14\Big[(D-1)(b_2+b_3)^2-4b_1^2\Big]A_\mu A_\nu+\frac{b_1}{2}\Big[\Big(2b_1-(D-1)(b_2+b_3 )\Big)A^2-2\divA \Big] g_{\mu\nu}\nonumber\\
&&+\Big[b_1+b_3-\frac D2(b_2+b_3)\Big]F_{\mu\nu}+\frac12\Big[2b_1-(D-1)(b_2+b_3)\Big] \nabla_\nu A_\mu\; ,\\
\mQ_{\mu\nu}&=&\frac12\Big[2b_1-(D+1)b_2+(D-1)b_3\Big]F_{\mu\nu}\; ,\\
\mP_{\mu\nu}&=&-R_{\mu\nu}+\frac14\Big[(b_2+b_3)^2-4(D-1)b_1^2\Big]A_\mu A_\nu-\frac{b_2+b_3}{4}\Big[\Big(b_2+b_3 -2(D-1)b_1\Big)A^2-2\divA \Big] g_{\mu\nu}\nonumber\\
&&+\Big[(D-1)b_1-b_2\Big]F_{\mu\nu}+\frac12\Big[2(D-1)b_1-b_2-b_3)\Big] \nabla_\nu A_\mu\;,
\label{RPQ}
\eea
where $R_{\mu\nu}$ and $\nabla$ are the Ricci tensor and covariant derivative of the Levi-Civita connection of the spacetime metric, $F_{\mu\nu}=\partial_\mu A_\nu-\partial_\nu A_\mu$ is the strength tensor of the vector field and $D$ is the spacetime dimension. Notice that the Ricci tensor $\mR_{\mu\nu}$ is not symmetric, not even in the torsion free case with $b_3=0$, since non-metricity can also induce an antisymmetric part for the Ricci tensor. It is also convenient to keep in mind that the homothetic tensor is always antisymmetric and for symmetric connections it is proportional to the antisymmetric part of the Ricci tensor. Finally, the Ricci scalar is unambiguously defined as $\mR=g^{\mu\nu}\mR_{\mu\nu}=-\mP_\mu{}^\mu$ and, for our connection (\ref{nonmetricity2}), it is given by
\be
\mR=R+\frac{D-1}{4}\Big[4b_1^2+(b_2+b_3)^2-2b_1(b_2+b_3)D\Big] A^2-\frac{D-1}{2}(2b_1+b_2+b_3)\divA\,,
\label{eqmR}
\ee
where $R$ is the Ricci scalar of the Levi-Civita connection. 

To end this section we will give some important geometrical objects. A defining property of geometries with non-metricity is that the length of vectors is not preserved under parallel transport. If we have a vector $v^\alpha$, its length $v^2=g_{\alpha\beta}v^\alpha v^\beta$ under a parallel displacement $\diff x^\mu$ is 
\be
\diff v^2=Q_{\mu\alpha\beta}v^\alpha v^\beta \diff x^\mu=\Big[(b_3-b_2)A_\mu v^2+(2b_1-b_2-b_3)A_\alpha g_{\beta\mu}v^\alpha v^\beta\Big]\diff x^\mu.
\ee
Obviously, for a pure metric geometry with $A_\mu=0$, the length is conserved. However, it is remarkable that the presence of torsion ($b_3$) also allows to have a wider class of geometries that preserve the length of vectors given by $b_3=b_2=b_1$. This actually allows to avoid one of the main problems of the pure Weyl geometries, where the length of a vector changes as it is parallel transported and, consequently, the properties of a physical object may depend on its history\footnote{It is important to emphasize that this crucially depends on the connection seen by matter fields. For instance, if bosonic fields are minimally coupled to the geometry, then they will be sensitive only to the Levi-Civita piece of the connection.}. In our more general geometrical set-up, this can be avoided even for non-trivial connections with non-metricity. One could relax the condition of invariance of the length of vectors only when they are transported along closed loops. In that case, the variation is determined by the homothetic or segmental curvature tensor $\mathcal{Q}_{\mu\nu}$ introduced above. Again, we have a special family of geometries with $2b_1-(D+1)b_2+(D-1)b_3=0$ in which the length of a vector remains invariant when transported around a closed path. Of course, these geometries contain the aforementioned case with $b_3=b_2=b_1$, but more general cases are possible in which the length of a vector may vary under a parallel transport while remaining the same when the trajectory closes.

Besides the variation of the length of vectors, the presence of non-metricity will also affect the properties of the geodesics. Since only the symmetric part of the connection enters the geodesic equation, the torsion term determined by $b_3$ will not enter here. A crucial feature of the geodesics is their projective similarity, i.e., two families of geodesics related by a change of affine parameter will describe the same class of paths. More precisely, it can be shown that two connections differing by $\delta^\alpha_{(\mu}\xi_{\nu)}$ give rise to the same geodesics up to a redefinition of the affine parameter \cite{Thomasprojective}. The projective invariant object determining the class of paths is the Thomas projective parameter, which for our connection is given by:
\be
\hat{\Pi}^\alpha_{\mu\nu}\equiv\hGam^\alpha_{\mu\nu}-\frac{2}{D+1}\hGam^\lambda_{\lambda(\mu}\delta^\alpha{}_{\nu)}=\Pi^\alpha_{\mu\nu}+\frac{b_1}{D+1}\Big[A_\mu\delta_\nu^\alpha+A_\nu\delta_\mu^\alpha-A^\alpha g_{\mu\nu} \Big]\,,
\ee
where $\Pi^\alpha_{\mu\nu}$ is the piece corresponding to the Levi-Civita part of the connection. In the above expression, only the symmetric (torsion-free) part of the connection should be considered. As expected, this expression does not depend on $b_2$, since this term precisely corresponds to a projective transformation and, consequently, the geodesic trajectories will not be affected by it.

\subsection{Gravitational actions}

Theories based on the Ricci scalar given in (\ref{eqmR}) lead to interesting phenomenologies, including the Starobinsky inflationary model and its so-called $\alpha$-attractor generalisation \cite{Ozkan:2015iva,Ozkan:2015kma,Jimenez:2015fva}. This is possible because in $f(\mR)$ type of theories with (\ref{eqmR}) the vector field is dynamically constrained to be the gradient of a scalar and, thus, similarly to usual $f(R)$ theories, they are equivalent to a scalar-tensor theory.

In order to have a fully dynamical vector field (and not imposed to be a pure gradient) we need to consider more general actions. The natural step is then to include quadratic curvature terms. For Levi-Civita connections, the requirement of having second order equations of motions (so that Ostrogradski instabilities are avoided) leads to the well-known Lovelock invariants. The quadratic term of such invariants is the Gauss-Bonnet term, which in 4 dimensions is a total derivative and, thus, it does not contribute to the gravitational field equations.  Considering the corresponding Gauss-Bonnet term with our vector connection will also give a total derivative in 4 dimensions. However, as found in \cite{Jimenez:2014rna}, more general quadratic curvature terms can give rise to a new interesting class of vector-tensor theories in the context of Weyl geometry, see also \cite{Haghani:2014zra}. Following the same procedure, we will write down the most general action quadratic in curvature invariants in the more general linear vector geometry defined by the connection (\ref{nonmetricity2}). Such terms are given by
 \begin{align}
\mS_{\rm quadratic}  = &\mu \int \diff^D x \sqrt{-g}\Big[\mR^2 + \mR_{\al\bt\gamma\delta}\Big( d_1 \mR^{\al\bt\gamma\delta} + d_2 \mR^{\gamma\delta\al\bt} 
-  d_3 \mR^{\al\bt\delta\gamma}\Big)\nonumber\\
& -  4\Big( c_1 \mR_{\mu\nu}\mR^{\mu\nu} +  c_2 \mR_{\mu\nu}\mR^{\nu\mu} 
+  \mP_{\mu\nu}\lp c_3 \mP^{\mu\nu} + c_4 \mP^{\nu\mu} -  c_5 \mR^{\mu\nu} - c_6 \mR^{\nu\mu}\rp 
+  \mQ_{\mu\nu}(c_7 \mQ^{\mu\nu} + c_8 \mR^{\mu\nu}+ c_9\mP^{\mu\nu})\Big)\,   
\Big]\,.
\label{Squadratic}
\end{align}
where $c_{i}$ and $b_i$ are dimensionless constants and $\mu$ has dimension $[{\rm mass}]^{4-D}$.  In order not to have Ostrogradski instabilities associated to higher order equations of motion for the metric we will restrict the parameters in order to recover the Gauss-Bonnet term for the Levi-Civita part of the connection (i.e, when $A_\mu=0$) so that
\be
d_1+d_2+d_3=\sum_{i=1}^6c_i=1\,.
\label{GBcondition}
\ee
Since $\mQ_{\mu\nu}$ identically vanishes for $A_\mu=0$, the coefficients $c_7,c_8,c_9$ remain fully free. Now we can rewrite (\ref{Squadratic}) as a vector-tensor theory in a Riemannian geometry. To that end, we will use the decomposition of the connection (\ref{nonmetricity2}) in (\ref{Squadratic}) and express everything in terms of $A_\mu$ and the Levi-Civita connection $\Gamma$. After some straightforward algebra and a few integrations by parts as done in \cite{Jimenez:2014rna}, the action can finally be expressed as 
\begin{align}
\mS_{\rm quadratic}=\mu\int\diff^Dx\sqrt{-g}&\left[\Big(R^2-4R_{\mu\nu}R^{\mu\nu}+R_{\mu\nu\rho\sigma}R^{\mu\nu\rho\sigma}\Big)-\frac\alpha 4 F_{\mu\nu} F^{\mu\nu}+\xi A^2\divA -\lambda A^4-\beta G^{\mu\nu} A_\mu A_\nu\right.\nonumber\\&
\gamma_1(\divA)^2+\big(\gamma_2 A^2+\gamma_3\divA\big)R\Big]\,,
\end{align}
where $\alpha$, $\beta$, $\gamma_i$, $\lambda$ and $\xi$ are dimensionless parameters that depend on $b_i$, $c_i$, $d_i$ and the spacetime dimension $D$. The combination in the brackets in the first line is nothing but the Gauss-Bonnet term for the Levi-Civita connection, as a consequence of having imposed (\ref{GBcondition}). The remaining terms in the first line are the same as were obtained in \cite{Jimenez:2014rna} for the case of pure Weyl geometry. Despite the derivative interaction $\xi$ and the non-minimal coupling $\beta$, those terms only propagate 3 degrees of freedom, very much like the simpler case of a Proca field.  However, the terms in the second line were not present in the pure Weyl case and only arise for our general connection (\ref{nonmetricity2}). These are however undesirable because they will 
propagate one additional degree of freedom besides the 3 polarizations corresponding to a massive vector field and this extra mode will generically suffer from the Ostrogradski instability, i.e., it will be a ghost. Thus, in order to have a stable theory we need to impose the conditions $\gamma_i=0$. The details are given in the Appendix A. Remarkably, the only solution for $D\geq4$ is $b_3=2b_1-b_2$, which exactly coincides with the generalised Weyl geometry discussed above that preserves the local conformal invariance of the metric compatibility condition. In that case, we can canonically normalize the field by means of the rescaling $A_\mu\rightarrow A_\mu/\sqrt{\alpha\mu}$. We can include the standard Einstein-Hilbert term $-\frac12 \mpl^{D-2}\mR$ for completeness, which will simply give a mass term for the vector field, so the final action for the vector reads
\begin{align}
\mS=&\int\diff^Dx\sqrt{-g}\left[-\frac1 4 F_{\mu\nu} F^{\mu\nu}+\frac12M^2 A^2+\xi A^2\divA -\lambda A^4-\beta G^{\mu\nu} A_\mu A_\nu\right]\,,
\label{TheAction}
\end{align}
with
\begin{eqnarray}
%M^2&\equiv&-\frac{D-1}{4}\Big[4b_1^2+(b_2+b_3)^2-2b_1(b_2+b_3)D\Big]\frac{ M_p^{D-2}}{\alpha\mu}\,,\\
M^2&\equiv&b_1^2(D-2)(D-1)\frac{ \mpl^{D-2}}{\alpha\mu}\,,\\
\xi&\equiv&4b_ 1^3\frac{(D-4)(D-3)(D-2)}{\alpha}(\alpha\mu)^{-1/2}\,,\\
\lambda&\equiv&-b_1^4\frac{(D-4)(D-3)(D-2)(D-1)}{\alpha}(\alpha\mu)^{-1}\,,\\
\beta&\equiv&-4b_1^2\frac{(D-4)(D-3)}{\alpha}\,.   
\end{eqnarray}
As we see, in 4 spacetime dimensions only the mass term remains while all the interactions vanish. To obtain a non-trivial theory in 4 dimensions we could consider the limit $D\rightarrow 4$ simultaneously with $\alpha\rightarrow 0$ while keeping $(D-4)/\alpha$ and $\alpha\mu$ fixed. Then there remains only two free parameters, in particular the action could be reduced to (when choose the free parameters as the constants $\xi$ and $\beta$)
\be \label{ReducedAction}
\mS \rightarrow \int\diff^4 x\sqrt{-g}\left[-\frac1 4 F_{\mu\nu} F^{\mu\nu}+ \frac{3}{4}\lp \frac{\xi}{\beta}\mpl\rp^2 A^2+\xi A^2\divA - \frac{3 \xi^2}{8\beta} A^4-\beta G^{\mu\nu} A_\mu A_\nu\right]\,.
\ee
Throughout this work, we will focus on $D=4$ and for generality consider all 4 parameters as independent.  
It can be useful to note that our general vector-tensor action can be rewritten in an alternative way by using that $G^{\mu \nu}A_\mu A_\nu=R^{\mu \nu}A_\mu A_\nu-\frac12 R A^2$ together with $R^{\mu \nu}A_\mu A_\nu=A^\nu[\nabla_\mu,\nabla_\nu]A^\mu$ so that we finally obtain
\be
\mS=\int\diff^4x\sqrt{-g}\left[-\frac{1}{2}\mpl^2\left(1-\frac{\beta A^2}{\mpl^2}\right) R-\frac{1+2\beta}{4} F_{\mu\nu} F^{\mu\nu}+\xi A^2\nabla_\mu A^\mu+\beta\Big[(\nabla_\mu A_\nu)^2-(\nabla_\mu A^\mu)^2\Big] +\frac12 M^2A^2-\lambda A^4\right]\,,
\ee
where we can recognize the typical Horndeski form for the vector field derivative self-interactions.

So far we have only considered actions up to quadratic order in curvature invariants. The same program can be straightforwardly applied to higher order curvature invariants by imposing the terms at a given order to reduce to the corresponding Lovelock invariant. Instead of discussing the general framework for this construction, we will simply comment on a class of terms that give rise to additional non-trivial derivative interactions for the vector field. For simplicity, we will consider the case $D=4$. At cubic order we know that a healthy coupling is given by the Horndeski vector-tensor interaction \cite{Horndeski:1976gi} $L^{\mu\nu\alpha\beta} F_{\mu\nu} F_{\alpha\beta}$, with $L^{\mu\nu\alpha\beta}=-\frac12 \epsilon^{\ \mu\nu\rho\sigma}\epsilon^{\alpha\beta\gamma\delta} R_{\rho\sigma\gamma\delta}$ the double dual Riemann tensor,  so we can use it to construct new terms in our class of geometries. For that, we notice that the antisymmetric part of the 3 independent traces of the Riemann tensor given in (\ref{RPQ}) are all proportional to $F_{\mu\nu}$. Thus, we can consider an interaction of the form 
\be
-\frac12 \epsilon^{\mu\nu\rho\sigma}\epsilon^{\alpha\beta\gamma\delta} \mR_{\rho\sigma\gamma\delta}\Big(f_1\mR_{[\mu\nu]}\mR_{[\alpha\beta]}+f_2\mR_{[\mu\nu]}\mP_{[\alpha\beta]}+f_3\mR_{[\mu\nu]}\mQ_{[\alpha\beta]}+f_4\mP_{[\mu\nu]}\mP_{[\alpha\beta]}+f_5\mP_{[\mu\nu]}\mQ_{[\alpha\beta]}+f_6\mQ_{\mu\nu}\mQ_{\alpha\beta}\Big)\,,
\ee
that will give the Horndeski interaction for the non-minimal derivative coupling of the vector. Although we have written down explicitly all the possible terms, they all will contribute exactly the same interactions for the vector field. Explicitly, the above cubic interaction leads to 
\begin{align}
\mS_{\rm cubic}=\mu_3^{-2}\int\diff^4x\sqrt{-g}\Big[&L^{\mu\nu\alpha\beta} F_{\mu\nu} F_{\alpha\beta}+2(2b_1+b_2+b_3)\Fd^{\mu\alpha}\Fd^\nu{}_\alpha\nabla_\mu A_\nu\nonumber\\
&+\frac{1}{2}\Big[\Big(2b_1-b_2-b_3\Big)^2A^2g^{\mu\nu}-2\Big(4b_1^2+(b_2+b_3)^2\Big)A^\mu A^\nu\Big]F_{\mu\alpha} F_\nu{}^\alpha\Big]\,,
\end{align}
where $\Fd^{\mu\nu}=\frac12 \epsilon^{\mu\nu\alpha\beta}F_{\alpha\beta}$ is the dual of the vector field strength tensor and $\mu_3$ some energy scale. The first term is the advertised Horndeski vector-tensor interaction (which has been studied in, e.g.,  \cite{Barrow:2012ay,Jimenez:2013qsa}), while the second term in the first line is a derivative self-interaction of the vector field. Although this term does not respect gauge invariance and contains derivatives of the vector field, its structure does not spoil the constraint making $A_0$ non-dynamical. This can be seen by noticing that the time derivative of $A_0$ couples to $\Fd^{0i}\Fd^0{}_ i$, which is proportional to the magnetic part of $F_{\mu\nu}$. Since this magnetic component only contains spatial gradients, $A_0$ will not acquire second derivatives in the field equations and, consequently, it will not propagate.  These interactions are in fact within the class of derivative self-interactions for a massive vector field discussed in \cite{Jimenez:2016isa,Allys:2015sht,Tasinato:2014eka,Heisenberg:2014rta}. Moreover, it is expected that by considering higher order curvature terms within our framework, the higher order terms interactions introduced in these references will be generated.

\section{de Sitter solutions}
\label{ds}

In this section we will show the existence of isotropic de Sitter solutions for the vector-tensor theory given by (\ref{TheAction}). In order to comply with the given symmetries, we will consider a purely temporal and homogeneous vector field $A_{\mu}=A_0(t)\delta_\mu{}^0$ configuration as well as a homogeneous and isotropic metric described by the Friedmann-Lema\^itre-Robertson-Walker (FLRW) line element
\be
\diff s^2=\diff t^2-a^2(t)\diff \vec{x}^2\,,
\ee
where $a(t)$ is the scale factor. The only non-trivial vector field equation for this configuration is given by
\be
A_0\left(M^2-4\lambda A_0^2+6\xi A_0 H-6\beta H^2\right)=0\,,
\ee
with $H=\dot{a}/a$ the Hubble expansion rate. As expected, this is an algebraic equation showing that $A_0$ is not dynamical and it is fully determined by $H$. This equation leads to 3 branches, namely: the trivial one with $A_0=0$ and 2 non-trivial ones with
\be
A_0=\frac{3\xi H\pm\sqrt{4\lambda M^2+(9\xi^2-24\lambda\beta)H^2}}{4\lambda}\,.
\label{A0solution}
\ee
From this expression we can see that de Sitter solutions are not guaranteed for any values of the parameters. For the singular case $\lambda=0$, the degree of the equation is reduced and only one non-trivial solution remains. This particular case will be studied separately below. In order to obtain the value of the Hubble parameter for the de Sitter solution we need to look at the corresponding Friedmann equation
\bea
3\mpl^2H^2=\frac{A_0^2}{2}\left(M^2-6\lambda A_0^2+12\xi A_0 H-18\beta H^2\right).
\eea
For the trivial branch with $A_0=0$ we obtain that $H=0$, i.e., we recover the Minkowski solution as expected. On the other hand, for the non-trivial branch given by (\ref{A0solution}), the Friedmann equation gives an algebraic equation for $H$. Such an equation forces $H$ to be constant and, therefore, these branches actually correspond to a de Sitter universe. However, the existence of such solutions will be subject to the existence of real solutions for the corresponding system of algebraic equations, which imposes restrictions on the parameters as already mentioned above. In fact, we can see that the equations reduce to a system of two polynomial (one quadratic and one quartic) equations so that we will have in general up to 8 different branches. They have some properties that can simplify the analysis of the solutions. First, one can easily see that there is a symmetry in the equations $(A_0,H)\rightarrow (-A_0,-H)$. Moreover, if $H$ is a solution for $\xi$, then $-H$ will be a solution for $-\xi$. Thus, without loss of generality we can absorb $\xi$ into a rescaling of $H\rightarrow H/\xi$ and focus on solutions with $H>0$ (i.e., expanding solutions), keeping in mind that a corresponding contracting solution will be guaranteed to exist as well. Further, one can get rid of another parameter (up to its sign) by absorbing it into the normalization of $A_0$. The most convenient one is to absorb $\beta$ so that $A_0\rightarrow A_0/\sqrt{\vert \beta \vert}$.

\begin{figure}
\begin{center}
\includegraphics[width=8cm]{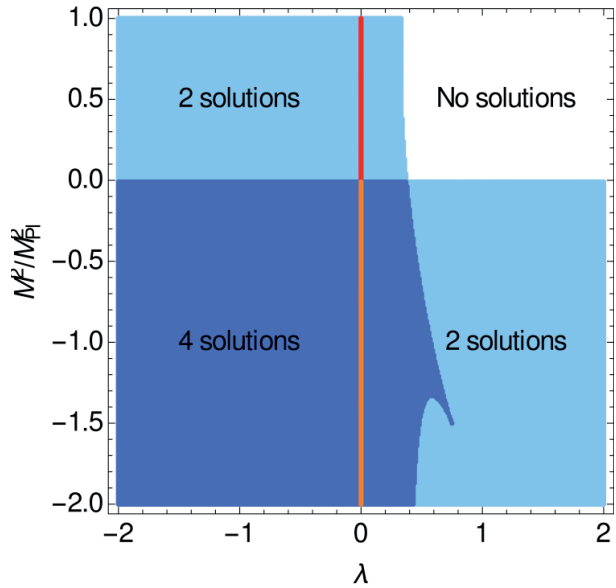}
\caption{In this plot we show the region in the parameter space where de Sitter solutions exist. Moreover, we also indicate how many de Sitter branches we find in each region. The red and orange vertical lines correspond to the singular case $\lambda=0$ where we find 1 and 3 solutions for $M^2$ positive and negative respectively.}
\label{Fig:deSittersolutions}
\end{center}
\end{figure}

In order to characterize the models for which de Sitter solutions exist, we will introduce the following rescalings and a dimensionless variable $x$
\be
M^2\rightarrow\frac{\xi^2}{\beta^2}M^2\,,\quad \lambda\rightarrow \frac{\xi^2}{\beta}\lambda\,,       \quad x\equiv \frac{\xi A_0}{\beta H}\,.
\ee
Obviously, this can only be done if both $\beta$ and $\xi$ are non-vanishing. From now on, we will assume this and treat separately below the case of vanishing $\beta$ or $\xi$. Then, from the vector field equation we can solve for $H$ in terms of $x$ as
\be
H^2=\frac{\xi^2 M^2}{\beta^3(6-6x+4\lambda x^2)}\,.
\label{solHx}
\ee
This already allows us to put some conditions on the parameters to have de Sitter solutions, since we need to have $H^2>0$. Now we can plug this solution into the Friedman equation to obtain the following quartic equation for $x$:
\be
18-18x+6\Big(2\lambda+\frac{M^2}{\mpl^2}\Big) x^2-3\frac{M^2}{\mpl^2} x^3+\lambda \frac{M^2}{\mpl^2}x^4=0\,.
\ee
This is the crucial equation that will determine the existence of de Sitter solutions. From here we see that if $\lambda M^2$ is negative, there will be at least a couple of real solutions. It is important to notice that, since this equation does not depend on $\beta$, we can take its real solutions, plug them into (\ref{solHx}) and choose the sign of $\beta$ in order to guarantee that $H^2$ is positive. In Fig. \ref{Fig:deSittersolutions} we show the number of solutions in the parameter space spanned by $(\lambda,M^2)$. For the singular case $\lambda=0$ it is not difficult to see that there is 1 solution for $M^2>0$ and 3 for $M^2<0$. 

We can see that, quite generally, the vector-tensor theories considered here give rise to de Sitter solutions. The next natural question is then if such solutions are stable. This is the subject of the next subsection.

\subsection*{Stability}

After showing the existence of de Sitter solutions, we will now proceed to check the stability of such solutions. For that, we will study the behaviour of the inhomogeneous perturbations around the de Sitter solutions found above. Thus, the background will be given by a constant temporal component for the vector field $A_0$ and a constant Hubble expansion rate $H$. The perturbations for the metric $g_{\mu\nu}$ will be decomposed into irreducible representations of the $SO(3)$ symmetry of the background in the usual manner \cite{Mukhanov:1990me}\begin{eqnarray}\label{perturbed_metric}
\delta g_{00} &=& -2\,\,\Phi\,,\nonumber\\
\delta g_{0i} &=& \,a\,\left(\partial_i B+B_i\right)\,,\nonumber\\
\delta g_{ij} &=& a^2 \left[2\,\delta_{ij}\psi +\left(\partial_i\partial_j-\frac{\delta_{ij}}{3}\partial^k\partial_k\right)E+\partial_{(i}E_{j)}+h_{ij}\right]\,,
\end{eqnarray}
where it is understood that all the metric perturbations depend on time and space and we have the constraints $\delta^{ij}h_{ij} = \partial^ih_{ij} = \partial^i E_i = \partial^i B_i=0$.  On the other hand, the vector field will be analogously perturbed as
\begin{eqnarray}
\dA_\mu=\Big(\dA_0,\partial_i A_s+\dA_i\Big) \,,
\end{eqnarray}
with $\delta^{ij}\partial_ i\dA_j=0$.

Na\"ively counted, we encounter 14 dof's in this decomposition, namely: $2\times1$ in the traceless symmetric tensor ($h_{ij}$), $2\times3$ in the divergence-free vectors ($B_i$, $E_i$, $\delta A_i$) and $1\times6$ in the scalars ($\Phi$, $B$, $\psi$, $E$, $\delta A_0$, $A_s$). However, we have 4 diffeomorphism gauge symmetries that remove 2 dof's each, so we have $14-2\times4=6$ dof's. In addition, as discussed above, the temporal component of the vector is not dynamical, but an auxiliary field, so we should substract yet another dof and, thus, the number of physical propagating modes will be $6-1=5$, i.e., the 2 polarizations corresponding to the massless graviton plus the 3 polarizations of the massive vector. In the following we will explicitly see this by studying the tensor, vector and scalar perturbations individually, and further establish the stability conditions for the 2 tensor, 2 vector and 1 scalar dof's in the theory.
Due to the homogeneity of the background, it will be convenient to decompose the perturbations in Fourier modes with respect to the spatial coordinates. Hence, all perturbations will be expanded as
\begin{equation}\label{tensor_Fourier}
\Theta(t,\vec{x})=\int \frac{\d^3k}{(2\pi)^{3/2}}\Big(\Theta_{\vec{k}}(t) e^{i\vec{k}\cdot\vec{x}} +\Theta^*_{\vec{k}}(t) e^{-i\vec{k}\cdot\vec{x}}\Big)\,,
\end{equation}
where $\Theta$ represents a given perturbation.

As a final remark, let us stress that we will perform the analysis without a priori fixing the gauge at the level of the action. We have also carried out the analysis in the Newtonian gauge where $B=E=B_i=0$  as a consistency check and found the same results.

\subsubsection*{Tensor perturbations}
 %h_{ij}=\int \frac{\d^3k}{(2\pi)^{3/2}}h_{ij,\vec{k}}(t) \exp(i\vec{k}\cdot\vec{x}) +c.c\,.
%\end{equation}
Let us start by studying the tensor perturbations. This is the simplest case because there are no dynamical dof's to be integrated out. After inserting the metric perturbations (\ref{perturbed_metric}) into our vector-tensor action with the fields decomposed in Fourier modes and using the dynamical background equations from the previous section, the action quadratic in the tensor perturbations reads 
\begin{equation}\label{action_tensormodes}
\mS^{(2)}_{\rm tensor} = \frac{\mpl^2}{8}\int \d^3k\,\d t\,\,a^3\,\left[\Big( 1+\frac{\beta A_0^2}{\mpl^2} \Big)\dot{h}_{ij,\vec{k}}^\star \dot{h}^{ij}_{\vec{k}}-\frac{k^2}{a^2}\left(1-\frac{\beta A_0^2}{\mpl^2}\right)h_{ij,\vec{k}}^\star h^{ij}_{\vec{k}} \right]\,.
\end{equation}
We see that the quadratic action for tensor perturbations is modified in the presence of a background $A_0$ and, as one would expect, only the non-minimal coupling to the Einstein tensor ($\beta$-term) contributes. However, there is a dependence on the remaining parameters through the background value $A_0^2$, which is determined by all the parameters.

From the above action, we easily conclude that tensor perturbations around the de Sitter solutions will avoid ghost-like instabilities, i.e., they have the right sign for the kinetic term, if we impose
\begin{equation}
 1+\frac{\beta A_0^2}{\mpl^2}  >0\,,
\end{equation}
which is trivially satisfied if $\beta\geq0$. On the other hand, the propagation speed of the perturbations is
\begin{equation}
c_t^2=\frac{ 1-\frac{\beta A_0^2}{\mpl^2}}{1+\frac{\beta A_0^2}{\mpl^2}}\,,
\end{equation}
that also must be positive to avoid gradient instabilities. These results are in agreement with those found in \cite{BeltranJimenez:2010uh}. We can avoid both ghosts and gradient instabilities for the tensor perturbations if
\be\label{condition_tensor}
\vert\beta\vert A_0^2< \mpl\,.
\ee

In any case, this non-trivial effect on the propagation speed of gravitational waves will be tightly constrained at present time since binary pulsar observations put stringent limits, at the level of $10^{-2}$ deviations from the speed of light \cite{Jimenez:2015bwa}.
%Concerning the three special examples that we have studied in detail in section \ref{examples_models}, as one can see from the stability condition of the tensor modes only $\beta$ and $A_0$ dictate their destiny. We shall look closer at each of the examples individually. Starting with the case $\lambda=0$ and $\xi=0$, we have that $H^2=M^2/(\beta6)$ and $A_0^2=\mpl^2 \beta/(1-3\beta^2)$. Therefore, the ghost and gradient instabilities for the tensor perturbations are avoided if $0<\beta<\sqrt{1/(3+\mpl)}$ or $\beta>1/\sqrt{3}$. On the other hand, for the case $\lambda=0$ and $\beta=0$, we have $A_0=-M^2/6\xi H$ and $H=M^{3/2}/(6^{3/4}\mpl^{1/2}\xi^{1/2})$ and the stability condition (\ref{condition_tensor}) is automatically fulfilled. Finally, for the case with $\lambda=0$ and $M^2=0$, we obtain that $A_0=\beta H/\xi$ and $H=\pm \mpl \xi/\sqrt{\vert\beta\vert}$ if $\beta$ is negative. But then the tensor perturbations would have ghost instabilities.

\subsubsection*{Vector perturbations}
Let us now turn to the slightly more involved case of vector modes. As we mentioned previously, two of the vector perturbations can be integrated out and only one vector will propagate (two dof's, as a transverse 3-vector). We expand our Lagrangian to second order in the vector perturbations and immediately observe that the vector field $B_i$ does not have any kinetic terms. Therefore we can simply compute the equation of motion with respect to $B_i$ and integrate them out. This yields
\begin{equation}
B_{i,\vec{k}}=\frac{2\beta A_0}{a(\mpl^2+\beta A_0^2)}\delta A_{i,\vec{k}}+\frac{1}{2}a\dot{E}_{i,\vec{k}}\,.
\end{equation}
After plugging this expression back into the action and adding total derivatives, the dependence on $E_i$ drops. Thus, as advertised, we end up with the quadratic action for only one vector,
\begin{equation}\label{action_vectormodes}
S^{(2)}_{\rm vector} = \frac{1}{2}\int \d^3k\,\d t\,a \left[
\delta\dot{A}_{i,\vec{k}}^\star \delta\dot{A}^i_{\vec{k}}-\frac{c_v^2 k^2}{a^2} \delta A_{i,\vec{k}}^\star \delta A^i_{\vec{k}} 
\right]\,.
\end{equation}
The propagation speed is given by
\begin{eqnarray}
c_v^2 &\equiv& \frac{ 1+\beta(1+2\beta) A_0^2/\mpl^2}{ 1+\beta A_0^2/\mpl^2}\,,
\end{eqnarray}
again in agreement with the findings in \cite{BeltranJimenez:2010uh}. As in the case of tensor perturbations, we see that only the coupling to the Einstein tensor ($\beta$-term) modifies the quadratic action of vector perturbations, and the remaining parameters only enter through the background value $A_0$.  We see that the vector perturbations are never ghostly and the only instability that can appear is a gradient one. This was expected because the kinetic term for our vector field is nothing but the usual Maxwell term. In order to avoid the Laplacian instability we have to require $c_v^2>0$. Interestingly, it is trivially satisfied for models with $\beta\geq0$.

%As next we would like to check the stability condition for the three different models studied in section \ref{examples_models}. Let us start again with the case $\lambda=0$ and $\xi=0$, where we had $A_0^2=\mpl^2 \beta/(1-3\beta^2)$ and hence $c_v^2=1+2\beta^3/(1-2\beta^2)$.  For the case $\lambda=0$ and $\beta=0$ we had $A_0=-M^2/6\xi H$ and $H=M^{3/2}/(6^{3/4}\mpl^{1/2}\xi^{1/2})$ and hence this yields $c_v^2=1$ and there is no gradient instability. On the other hand for the case with $\lambda=0$ and $M^2=0$, we had $A_0=\beta H/\xi$ and $H=\pm \mpl \xi/\sqrt{\vert\beta\vert}$ if $\beta$ is negative. Therefore, the sound speed is simply given by $c_v^2=1+\beta$. The gradient instability is avoided in this case if $-1<\beta<0$.
%More generally, for $\beta<0$ we need to demand
%\begin{eqnarray}\label{constraints_vector}
%&&\beta <0 \,, \qquad 1+2\beta>0 \qquad \text{and} \qquad \mpl^2+A_0^2\beta(1+2\beta)<0  \nonumber\\
%&&A_0^2>0  \qquad \text{and} \qquad \beta (\mpl+A_0^2\beta)<0  \nonumber\\
%&&A_0^2>0  \qquad \text{and} \qquad\beta>=0\,.
%\end{eqnarray}

\subsubsection*{Scalar perturbations}
As usual, the scalar sector is the most involved one. We have six scalars ($\psi$, $\delta A_0$, $A_s$, $E$, $B$, $\Phi$ ), but only one propagates, corresponding to the longitudinal mode of $A_\mu$. Again, we expand the action to quadratic order in the scalar perturbations. The first thing to be noticed is the fact that the corresponding kinetic matrix (or alternatively the Hessian matrix) contains already at this stage three vanishing eigenvalues, imposing three constraint equations that makes three out of the six scalar fields not propagating. The kinetic matrix is
\begin{eqnarray}
\mathcal{K}_{\psi, \delta A_0, A_s, E, B, \Phi}=
\begin{pmatrix}
-6(\mpl^2+\beta A_0^2)/\mpl^2&0&0&0 &0&0\\
0&0 &0&0 &0&0\\
0&0&k^2/(\mpl^2a^2) &0 &0&0\\
0&0&0&(\mpl^2+\beta A_0^2)/(6\mpl^2)&0&0 \\
0&0&0&0&0&0 \\
0&0&0&0&0&0
 \end{pmatrix}\,.
\end{eqnarray}
As one can see, the quadratic action does not evolve any kinetic term for the scalar fields $\Phi$ and $B$ and $\delta A_0$. Thus we can simply replace them by using their equations of motion. For instance, the equation of motion for the scalar field $\Phi$ gives
\begin{eqnarray}
\Phi_{\vec{k}}=\frac{12\beta H \delta A_0+\mpl^2(\dot{E}_{\vec{k}}+6\dot{\psi}_{\vec{k}})+A_0^2(-6\xi \delta A_0+\beta\dot{E}_{\vec{k}}+6\beta\dot{\psi}_{\vec{k}})}{-6\xi A_0^3+6(\mpl^2+3\beta A_0^2)H} \,.
\end{eqnarray} 
Similarly, the expressions for $\delta A_0$ and $B$ are obtained using their equations of motion, which we omit here. After inserting the solutions for $B$, $\Phi$ and $\delta A_0$ back into the quadratic action, the resulting expression depends only on the remaining three scalar fields ($\psi$, $A_s$, $E$). On a closer inspection, one realizes that the kinetic matrix of the three scalar fields still has a vanishing determinant, pointing to the presence of more constraints that can be used to integrate out some of the scalar fields. To be precise, the kinetic matrix contains two zero eigenvalues and one non-vanishing eigenvalue. 
%\begin{eqnarray}\label{3newvarMatrix}
%\mathcal{K}_{\psi, E, A_s}=
%\begin{pmatrix}
%0&0&0\\
%0&0 &0\\
%0&0&\lambda_3
% \end{pmatrix}
%\end{eqnarray}
%The corresponding eigenvectors to the three eigenvalues  take the form
%\begin{eqnarray}
%v_{1}=
%\begin{pmatrix}
% H/A_0\\
%0\\
%1
% \end{pmatrix}, \qquad  v_{2}=
%\begin{pmatrix}
%-k^2/6\\
%1\\
%0
% \end{pmatrix},  \qquad  v_{3}=
%\begin{pmatrix}
%-A_0/H\\
%-k^2A_0/(6H)\\
%1
% \end{pmatrix}
%\end{eqnarray}
%With the help of the rotational matrix built out of the eigenvectors 
%\begin{eqnarray}
%P=\begin{pmatrix}
%H/A_0&-k^2/6&-A_0/H\\
%0&1 &-k^2A_0/(6H)\\
%1&0&1
% \end{pmatrix}
%\end{eqnarray}
We can diagonalize the kinetic matrix by performing the following field redefinitions, which will make the only propagating scalar field manifest:
\begin{eqnarray}
F_{1,\vec{k}}&=&\frac{A_0(37A_0A_{s,\vec{k}}+6H( E_{\vec{k}}+6\psi_{\vec{k}}))}{37A_0^2+36H^2}\,, \nonumber\\
F_{2,\vec{k}}&=&\frac{6(A_0A_{s,\vec{k}}H+6E_{\vec{k}}H^2+A_0^2(6E_{\vec{k}}-\psi_{\vec{k}}))}{37A_0^2+36H^2}\,,\nonumber\\
F_{3,\vec{k}}&=&\frac{6H(6A_{s,\vec{k}}H-A_0(E_{\vec{k}}+6\psi_{\vec{k}}))}{37A_0^2+36H^2}\,.
\end{eqnarray}
After adding total derivatives, the resulting action depends only on $F_3$ and reads
\begin{equation}
S^{(2)}_{\rm scalar}= \frac{\mpl^2}{8}\int d^3k \,dt\,a^3 \,\left({\mathcal K}_s\dot{F}_{3,\vec{k}}\,\dot{F}_{3,\vec{k}}^* - {\mathcal V}_s\frac{k^2}{a^2}F_{3,\vec{k}}\,\,F_{3,\vec{k}}^*\right)\,,
\end{equation}
where ${\mathcal K}_s$ and ${\mathcal V}_s$ are some functions of the theory parameters and the background solution. Their exact expressions are very cumbersome in the general case. However, their UV limits with $k\rightarrow \infty$ can be expressed as 
\begin{align}
{\mathcal K}_{s,UV}&=-\frac{(A_0(\mpl^2+\beta A_0^2)(37A_0^2+36H^2)^2((4\beta \lambda-3\xi^2)A_0^3-3\mpl^2\xi H+9\beta \xi A_0^2H+4A_0(\mpl^2\lambda-3\beta^2H^2)))}{162(\mpl^2H^2(-\xi A_0^3+(\mpl^2+3\beta A_0^2)H)^2)}\, \label{kmatrix}\\
{\mathcal V}_{s,UV}&=\frac{A_0(2\mpl^2\xi A_0-\xi(1+2\beta)A_0^3+\mpl^2H+\beta(3+8\beta)A_0^2H)(37A_0^2+36H^2)^2(\mpl^2\xi-\beta A_0(\xi A_0-4\beta H))}{162\mpl^2a^2H^2(-\xi A_0^3+(\mpl^2+3\beta A_0^2)H)^2}k^2\,. \label{vmatrix}
\end{align}
For the stability of the perturbations we have to impose that both ${\mathcal K}_{s}$ and ${\mathcal V}_{s}$ be positive. Unlike in the previous cases, we see that for the scalar sector all the terms contribute to the perturbations and not only through the background evolution.

%Again, we shall investigate closer the stability conditions of the three special examples of section \ref{examples_models}. For the first case with $\lambda=0$ and $\xi=0$, we had that $H^2=M^2/(\beta6)$ and $A_0^2=\mpl^2 \beta/(1-3\beta^2)$. Therefore, the kinetic term in the UV becomes $\hat{K}_{UV}=4k^8\mpl^2\beta^6(1-2\beta^2)/(9(M-3M\beta^2)^2)$ whereas the gradient term becomes $\hat{M}_{UV}=4k^{10}\mpl^2\beta^6(1+8\beta^3)/(27M^2(1-3\beta^2)a^2)$. In order to avoid kinetic and gradient instabilities, one has to impose either $-1/2<\beta<0$, or $0<\beta<1/\sqrt{3}$ or $1/\sqrt{3}<\beta<1/\sqrt{2}$. For the second case with $\lambda=0$ and $\beta=0$, we had $A_0=-M^2/6\xi H$ and $H=M^{3/2}/(6^{3/4}\mpl^{1/2}\xi^{1/2})$. One immediately observes that for this case the kinetic term in the UV limit vanishes $\hat{K}_{UV}=0$. The scalar modes would become infinitely strongly coupled on this background. Finally, for the case with $\lambda=0$ and $M^2=0$, we had that $A_0=\beta H/\xi$ and $H=\pm \mpl \xi/\sqrt{\vert\beta\vert}$ if $\beta$ is negative. Hence the corresponding kinetic term in the UV simplifies to $\hat{K}_{UV}=k^8\beta^2/(81\xi^2)$ and $\hat{M}_{UV}=2k^10\beta^2(3+8\beta)/(729\xi^2a^2)$. In this case the scalar perturbations would be free of ghost and gradient instabilities if $-3/8<\beta<0$.

\section{Cosmological solutions}
\label{cosmology}

After showing the existence of de Sitter solutions and studying the corresponding perturbations around them, we will now consider a more general case in the presence of a matter component. Again, we will consider homogeneous and isotropic universes with the FLRW line element. In such a background metric and for a homogeneous vector field configuration with $A_{\mu}=(A_0(t),a^{-1}\vec{A}(t))$, the vector field equations read
\bea
A_0\Big(M^2+4\lambda\vec{A}^2-6\beta H^2+6\xi H A_0-4\lambda A_0^2\Big)+\xi\frac{\d \vec{A}^2}{\d t}=0\,,\\
\frac{\d^2\vec{A}}{\d t^2}+3H \frac{\d\vec{A}}{\d t}+\Big[M^2-4\lambda(A_0^2-\vec{A}^2)+2\xi(\dot{A}_0+3HA_0)-2\beta(2\dot{H}+3H^2)+2H^2+\dot{H}\Big]\vec{A}=0\,.
\eea
From these equations we see that it is a consistent Ansatz to consider purely isotropic solutions with $\vec{A}=0$ so that our isotropic solutions will be supported by the temporal component of the vector field. Having isotropic solutions based on the spatial part of the vector usually requires to have a set of vector fields with a global $SO(3)$ symmetry \cite{Bento:1992wy,ArmendarizPicon:2004pm}, sometimes called triad, or rapidly oscillating fields \cite{Cembranos:2012ng,Cembranos:2012kk}.

Let us now consider a universe filled with the vector field plus a matter component with energy density $\rho$. We will still restrict to purely isotropic solutions. Then, the vector field and Friedmann equations are
\bea
A_0\left(M^2-4\lambda A_0^2+6\xi A_0 H-6\beta H^2\right)=0\,,\label{eq:A0matter}\\
3\mpl^2H^2=\frac{A_0^2}{2}\left(M^2-6\lambda A_0^2+12\xi A_0 H-18\beta H^2\right)+\rho\,,\label{eq:Friedmannmatter}
\eea
respectively. For $A_0=0$ we have the trivial branch that recovers standard gravity so we will assume that $A_0\neq0$. In such a case, we can simply integrate out $A_0$ by (algebraically) solving its own equation of motion, whose solution is still given by (\ref{A0solution}) and, in general, it yields $A_0=A_0(H)$. Then, we can use this solution again in Friedmann equation to obtain a generalised version of it as
\be
3\mpl^2 H^2-\left[\frac{A_0^2}{2}\left(M^2-6\lambda A_0^2+12\xi A_0 H-18\beta H^2\right)\right]_{A_0=A_0(H)}=\rho\,.
\ee
Then, we can invert this equation to obtain a Cardassian-like model where the Hubble expansion rate is given by a non-linear function of the energy density. This is nothing but a particular example of the general result that gravity with auxiliary fields leads to a modified matter coupling. The explicit expression is
\bea
3\mpl^2H^2\left[\left(1+\frac{\beta M^2}{4\lambda\mpl^2}-3\frac{\xi^2M^2}{16\lambda^2\mpl^2}  \right) 
+3\epsilon\frac{\xi(8\beta\lambda-3\xi^2)}{32\lambda^3\mpl^2}H\sqrt{4\lambda M^2+(9\xi^2-24\lambda\beta)H^2}\right.\nonumber\\
\left.-\frac{9}{32\mpl^2}\left(\frac{3\xi^4}{\lambda^3}-\frac{12\beta\xi^2}{\lambda^2}+\frac{8\beta^2}{\lambda}\right)H^2
  \right]=\rho-\frac{M^4}{16\lambda}\,,
  \label{eq:FullFriedmann}
\eea
where $\epsilon=\pm 1$ parameterizes the two non-trivial branches. If we expand this expression for small $H$ we obtain
\be
3\mpl^2\left(1+\frac{4\beta\lambda-3\xi^2}{16\lambda^2}\frac{M^2}{\mpl^2}\right)H^2=\rho-\frac{M^4}{16\lambda}\,,
\label{eq:FriedmansmallH}
\ee
recovering the usual GR Friedmann equation with an effective cosmological constant and a rescaled Planck mass. Notice that the case with $\lambda=0$ is singular, as a consequence of reducing the degree of the vector field equation (which becomes linear in $A_0$ in that case). This singular case will be studied in more detail separately below. Also notice that the small-$H$ regime does not need to correspond to a low density regime.

A very interesting property of the cosmological evolution in the presence of the vector field is the possibility of having a maximum and/or minimum value for the Hubble expansion rate. This is determined by the discriminant of the vector field equation, i.e., the behavior of 
\be
4\lambda M^2+(9\xi^2-24\lambda\beta)H^2\,,
\ee
which must be positive in order to have real solutions. If both combinations $\lambda M^2$ and $9\xi^2-24\lambda\beta$ are positive, we always have real solutions, while if they are negative, real solutions do not exist. However, if they have different signs, we encounter two possibilities:
\begin{itemize}
\item If $\lambda M^2>0$ and $9\xi^2-24\lambda\beta<0$, there is an upper limit for $H$ given by
\be
H_{\star}=-\frac{4\lambda}{9\xi^2-24\lambda\beta}M^2\,.
\ee
\item If $\lambda M^2<0$ and $9\xi^2-24\lambda\beta>0$, then we have a lower bound for the Hubble expansion rate so that $H\geq H_\star$.
\end{itemize}

The above conditions guarantee that the solution for the vector field is real. However, we need additional conditions to guarantee the existence of physical solutions because, once the vector field has been solved for, we will obtain an equation for $H(t)$ which must also have real solutions. The overall effect will be the presence of bounds for the energy density of the matter component. The corresponding analysis for the general theory is very cumbersome, so we will focus on some particular cases instead below.

The models with an upper bound for $H$ could be useful to resolve the Big Bang singularities. This is expected to be a more general feature (not only for cosmology) that could help regularising other types of singularities, e.g. black hole singularities. Notice that this stems from the quadratic equation for the vector field.

\begin{figure}
\begin{center}
\includegraphics[width=8cm]{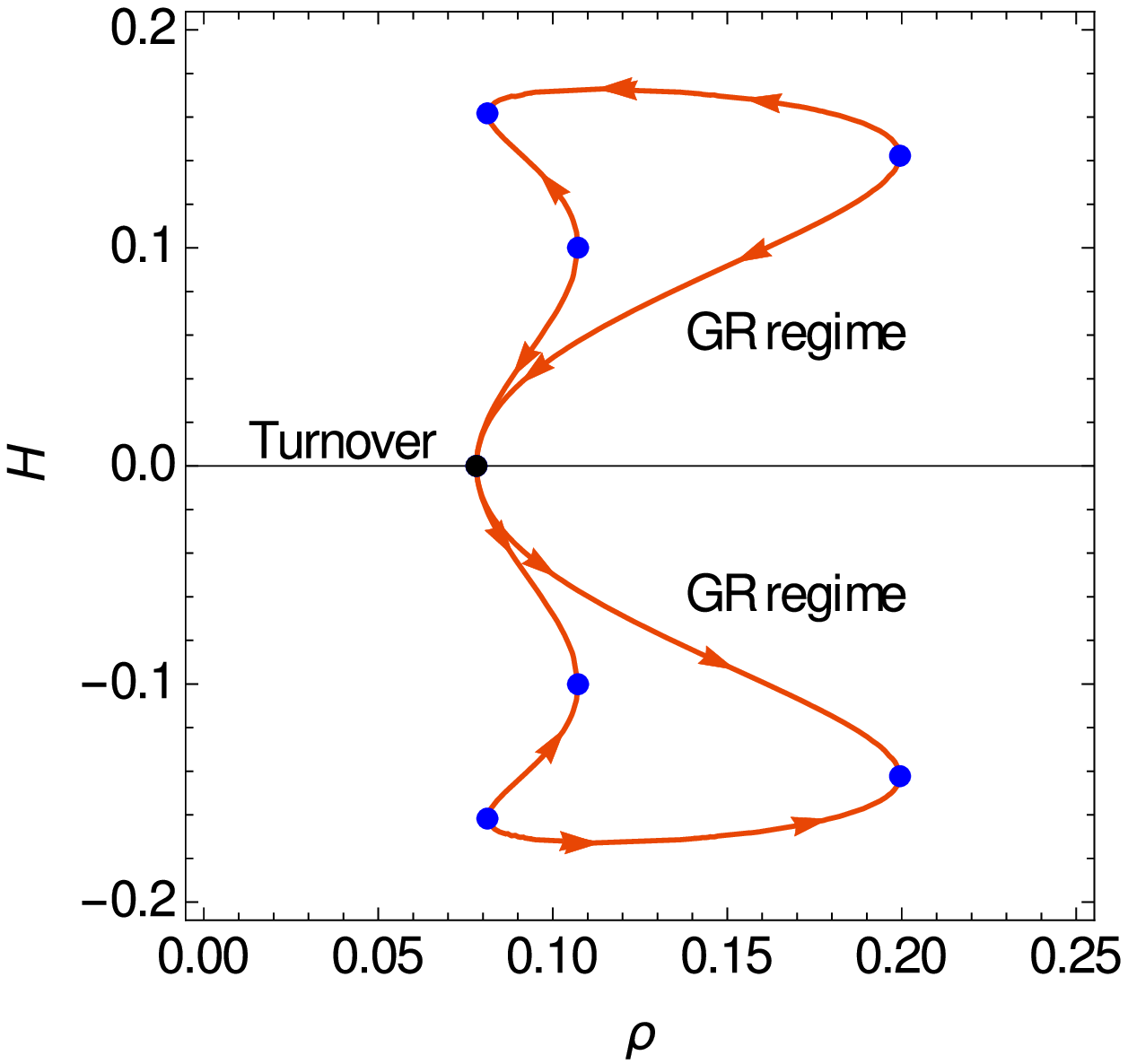}
\includegraphics[width=8cm]{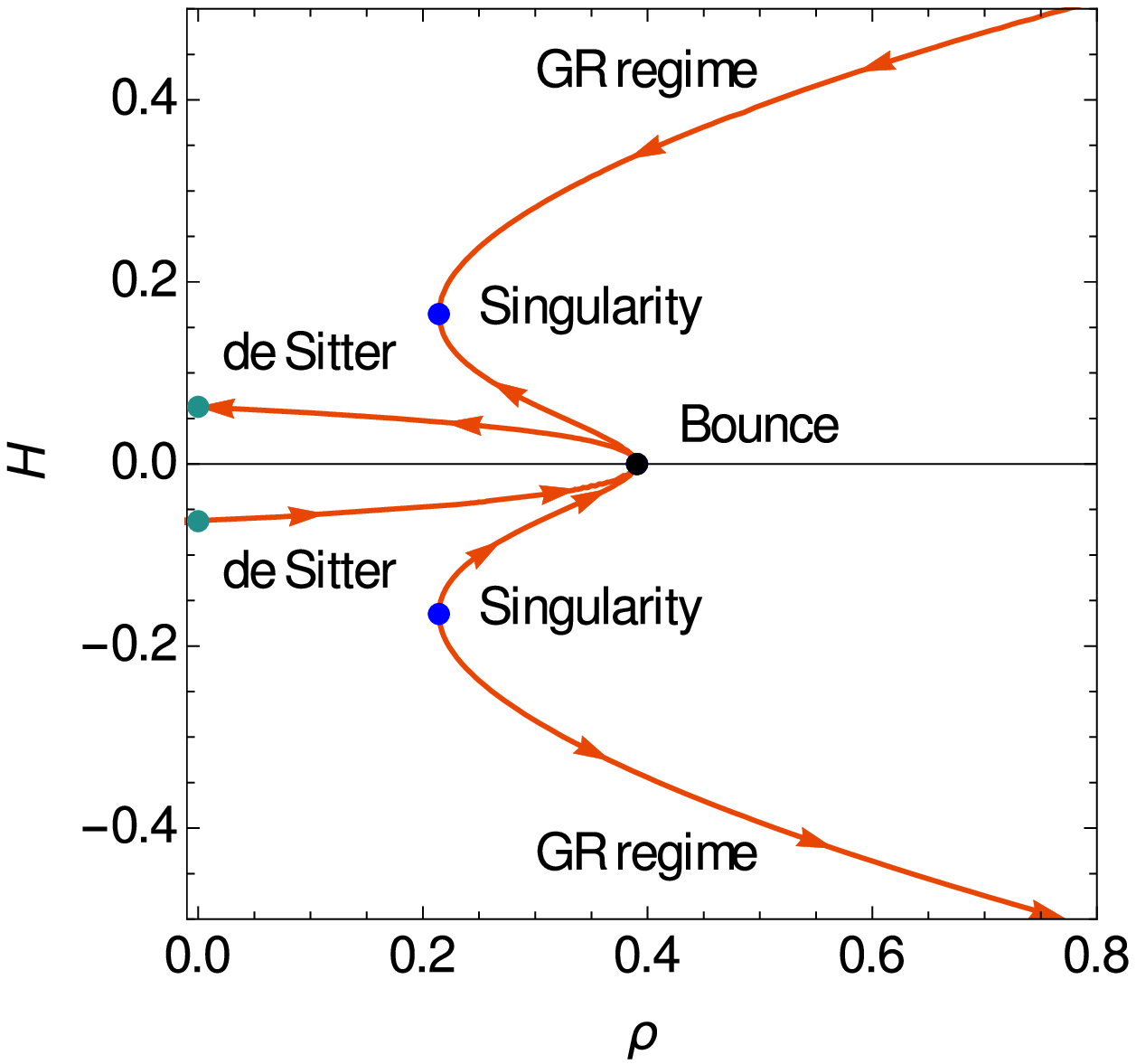}
\caption{ In this plot we show the solution of $H$ as a function of $\rho$ (normalized to Planck units) as obtained from Eq. (\ref{eq:FullFriedmann}) for two prototypical examples featuring the general properties described in detail in the main text. We have also indicated the flow under the time evolution. In the left panel ($\beta=3$, $\lambda=0.2$, $\xi=1/3$ and $M^2=0.5\mpl^2$) we can see 6 points (marked in blue) where we encounter a divergence in $\dot{H}$ and, thus, the evolution has a singularity. We also see a turnover matching an expanding universe with a contracting phase at low densities. Notice that at the turnover the different branches join. Also as discussed in the main text, we see that both $H$ and $\rho$ are bounded to a compact region. In the right panel ($\beta=0$, $\lambda=1$, $\xi=6$ and $M^2=2.5\mpl^2$) we can see the type of bounce that we can have in these theories. We see that from the bounce we can evolve either towards a low density de Sitter phase or end in a singularity. This illustrates the fact that a bounce cannot be connected with a GR regime at low densities and for small $H$. The GR regime in this case is beyond the singularities, in a region that is not continuously connected with the bounce.}
\label{Fig:Examples}
\end{center}
\end{figure}

\subsection{Bouncing solutions}

An interesting question that arises is whether it is possible to obtain viable bouncing solutions. As we have shown, the presence of the auxiliary field $A_0$ gives rise to a modified Friedmann equation and, therefore, it would be plausible to encounter some bouncing solutions. Such solutions are characterized by the existence of a finite (non-vanishing) energy density $\rho_b$ for which the Hubble expansion rate vanishes, i.e., $H(\rho_b)=0$ for $\rho_b\neq 0$. If we look at Eq. (\ref{eq:A0matter}), we see that for a potential bouncing solution we must have
\be
A_0^2=\frac{M^2}{4\lambda}\,,
\ee
so that it is only possible in models with $M^2$ and $\lambda$ different from zero and such that they have the same sign. This relation further implies that, at the bounce, the two branches of solutions for $A_0$ with $\epsilon=\pm1$ coincide. Thus, a generic behaviour will be the merging of branches at the bounce (this is explicitly illustrated in Fig \ref{Fig:Examples}). On the other hand, from the Friedmann equation we have that the potential bounce happens for an energy density
\be
\rho_b=\frac{M^4}{16\lambda}\,,
\ee
so that at the bounce we have the relation $\rho_b=4\lambda A_0^4$. Moreover, we see that $\lambda$ (and therefore $M^2$) must be positive in order to have a bounce with a positive density energy. However, for this bounce to actually describe a transition from contraction to expansion, we also need $\dot{H}$ to be positive near the bounce. This time derivative can be expressed as
\be
\dot{H}=-\frac{3}{2}(\rho+p)\frac{\diff H^2}{\diff \rho }\,,
\label{eq:Hdot}
\ee
where we have transformed the time derivative into derivative with respect to $\rho$ and used the continuity equation $\dot{\rho}=-3H(\rho+p)$. This expression shows the well-known fact that in GR we need to violate the null energy condition to have a bounce (since $\frac{\diff H^2}{\diff \rho }=1/(3\mpl^2)>0$ in that case). In our case, the derivative near the bounce can be easily computed from the Friedmann equation to give
\be
\frac{\diff H^2}{\diff \rho }\Big\vert_{H\rightarrow 0}=\frac{1}{3\mpl^2\left(1+\frac{\beta M^2}{4\lambda \mpl^2}-3\frac{\xi^2M^2}{16\lambda^2 \mpl^2}\right)} \,.
\ee
Thus, if we want to have a bounce corresponding to a transition from a contracting to an expanding phase for a matter component satisfying the null energy condition, we need
\be
1+\frac{\beta M^2}{4\lambda \mpl^2}-3\frac{\xi^2M^2}{16\lambda^2 \mpl^2}<0.
\ee
This condition implies that the effective Newton's constant in the Friedmann equation for small $H$ in Eq. (\ref{eq:FriedmansmallH}) is negative. Then, from that equation we see that the bouncing solution at high energy density must match a de Sitter universe at low densities. In particular, this precludes the possibility of recovering the usual GR Friedmann equation at low densities and, as a consequence, realistic bouncing scenarios will be difficult to realize.

On the other hand, as we showed above, the stability of tensor perturbations requires $1+\frac{\beta A_0^2}{ \mpl^2}>0$, which at the bounce precisely gives $1+\frac{\beta M^2}{4\lambda \mpl^2}$. Thus, we need to have $\xi\neq0$ for the bounce to be possible and avoid tensor instabilities. It is important to stress that these are necessary conditions, but not sufficient.

Another interesting scenario corresponds to the opposite case, i.e., a turnover in the cosmological evolution with a transition from an expanding to a contracting universe. This case corresponds to having $\dot{H}<0$ near the turnover so that we need to require
\be
1+\frac{\beta M^2}{4\lambda \mpl^2}-3\frac{\xi^2M^2}{16\lambda^2 \mpl^2}>0\,,
\ee
to guarantee the existence of these solutions. We can see that the transition from expanding to contracting universe is possible for models with $\xi=0$, unlike in the previous case where $\xi$ needed to be non-vanishing. Interestingly, this behaviour resembles the evolution in the presence of a non-vanishing curvature for the spatial sections, but for a flat FLRW metric. Thus, it is possible to mimic the effect of a non-vanishing curvature with the vector field.

\subsection{Singularities}

In the previous section we have studied the existence of turnonvers corresponding to transitions between contracting and expanding phases. However, we can also have another class of turnovers which actually lead to cosmological singularities. These cases correspond to points where $\diff H/\diff\rho$ becomes infinity for finite $H$ and $\rho$. If we consider again the expression (\ref{eq:Hdot}), we see that at those points there is a divergence in $\dot{H}$, while $H$ and $\rho$ remain finite. Since the divergence only appears in the derivative of the Hubble expansion rate, it is plausible that such a singularity does not lead to a singular spacetime, i.e., the spacetime could be geodesically complete. In other words, the geodesics might go smoothly through the singular point. On the other hand, another worrisome feature of having a singularity is that, even if the geodesics can go through, tidal forces might diverge so that extended objects can not go through the singularity (see e.g. \cite{Nojiri:2005sx,Jimenez:2016sgs} for further discussions about these points). We will not discuss in further detail the specific properties of these singularities here, but we will content ourselves with analyzing their attracting properties. The points where singularity occurs can be easily computed from Eq. (\ref{eq:FullFriedmann}) as those values of $H$ for which $\diff\rho/\diff H=0$. Then, near the turnover the Friedmann equation will read
\be
\rho-\rho_\star=\frac12\left(\frac{\diff^2\rho}{\diff H^2}\right)_\star \Big(H-H_\star\Big)^2\,,
\ee
where $\star$ denotes the value at the turnover. The value of $\left(\frac{\diff^2\rho}{\diff H^2}\right)_\star$ determines whether the function $\rho(H)$ has a minimum or a maximum at the turnover and, consequently, will determine whether it will be approached in the time evolution. It is important to remember that we are considering minimally coupled matter so that if $H>0$, the expansion of the universe implies that $\rho$ decreases and, analogously, $\rho$ increases in the evolution for $H<0$. Thus, if $H>0$ and $\left(\frac{\diff^2\rho}{\diff H^2}\right)_\star>0$, the turnover will be approached, while for $\left(\frac{\diff^2\rho}{\diff H^2}\right)_\star<0$ the universe will evolve away from it. On the other hand, if $H<0$ we have that the turnover is an attractor for $\left(\frac{\diff^2\rho}{\diff H^2}\right)_\star<0$ and a repeller for $\left(\frac{\diff^2\rho}{\diff H^2}\right)_\star>0$. These behaviours are illustrated in Fig. \ref{Fig:Examples}.

\subsection{Self-tuning solutions}
In this subsection we will investigate whether the vector-tensor theories that we are considering can lead to self-tuning solutions, i.e., Minkowski solutions in the presence of an arbitrary cosmological constant. For that we need to have $H=0$ for $\rho\neq 0$. The idea behind the self-tuning mechanism is to make the vector field equation trivial for $A_0$ in Minkowski so that, then, it can be used in the Friedmann equation to screen an arbitrary cosmological constant. In our case, making the equation of $A_0$ in Minkowski is only possible if $M^2=\lambda=0$. However, for that case all the dependence on $A_0$ from Friedmann equation also drops in a Minkowski solution. This means that it is not possible to realize self-tuning solutions. This is not surprising and, in fact, in the original proposal within the class of Horndeski theories considered in \cite{Charmousis:2011bf,Charmousis:2011ea} the presence of non-flat FLRW was crucial. In our case, we can show that the same conclusion can be reached. Let us consider a non-flat FLRW metric
\be
\diff s^2=\diff t^2-a^2(t)\left(\frac{\diff r^2}{1-kr^2}+r^2\diff\Omega^2\right)\,,
\ee
with $k$ the curvature of the spacial sections. Then, instead of imposing $H=0$ as it corresponds to pure Minkowski space, we impose the Ricci flat condition $H^2+k/a^2=0$. The vector field equation then reads
\be
A_0\left[M^2+6\xi\frac{\sqrt{-k}}{a}A_0-4\lambda A_0^2\right]=0\,.
\ee
This equation is trivial only if $M^2=\xi=\lambda=0$. Then, the Friedmann equation reduces to
\be
6\beta\frac{k}{a^2}A_0^2+\rho_\Lambda=0\,,
\ee
and we see that $A_0$ can dynamically compensate for the cosmological constant analogously to the original proposal. Of course, this does not come completely as a surprise since the theory with all the parameters vanishing but $\beta$ resembles one of the terms identified in \cite{Charmousis:2011bf,Charmousis:2011ea} to generate self-tuning solutions.

\section{Models examples}\label{examples_models}
Now that we have studied some general properties of the theory under consideration, in this section we will explore some specific models where the general features discussed above can be illustrated by simple cases which can be treated analytically. Among the considered examples, we will study the singular cases mentioned above.

\subsection{Case $\lambda=\xi=0$: de Sitter self-tuning}
\begin{figure}
\begin{center}
\includegraphics[width=8cm]{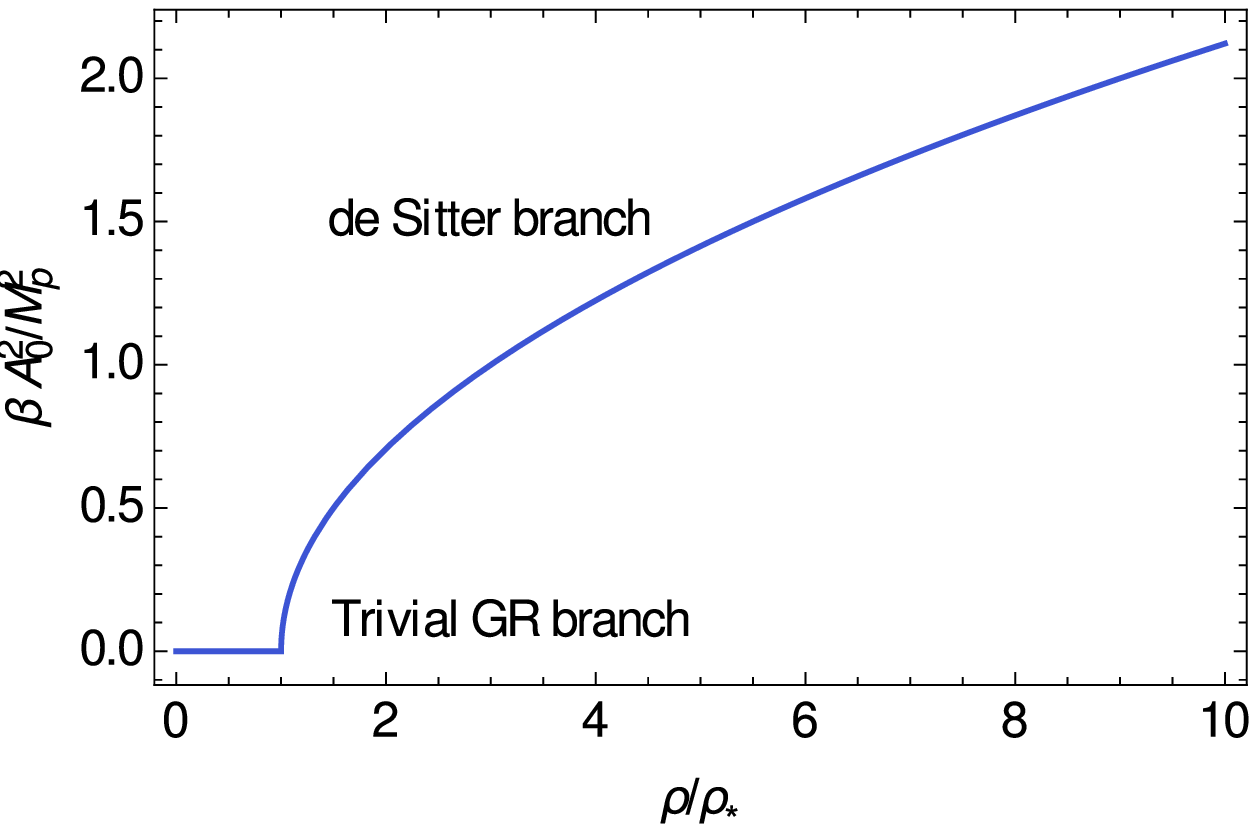}\hspace{0.75cm}
\includegraphics[width=8cm]{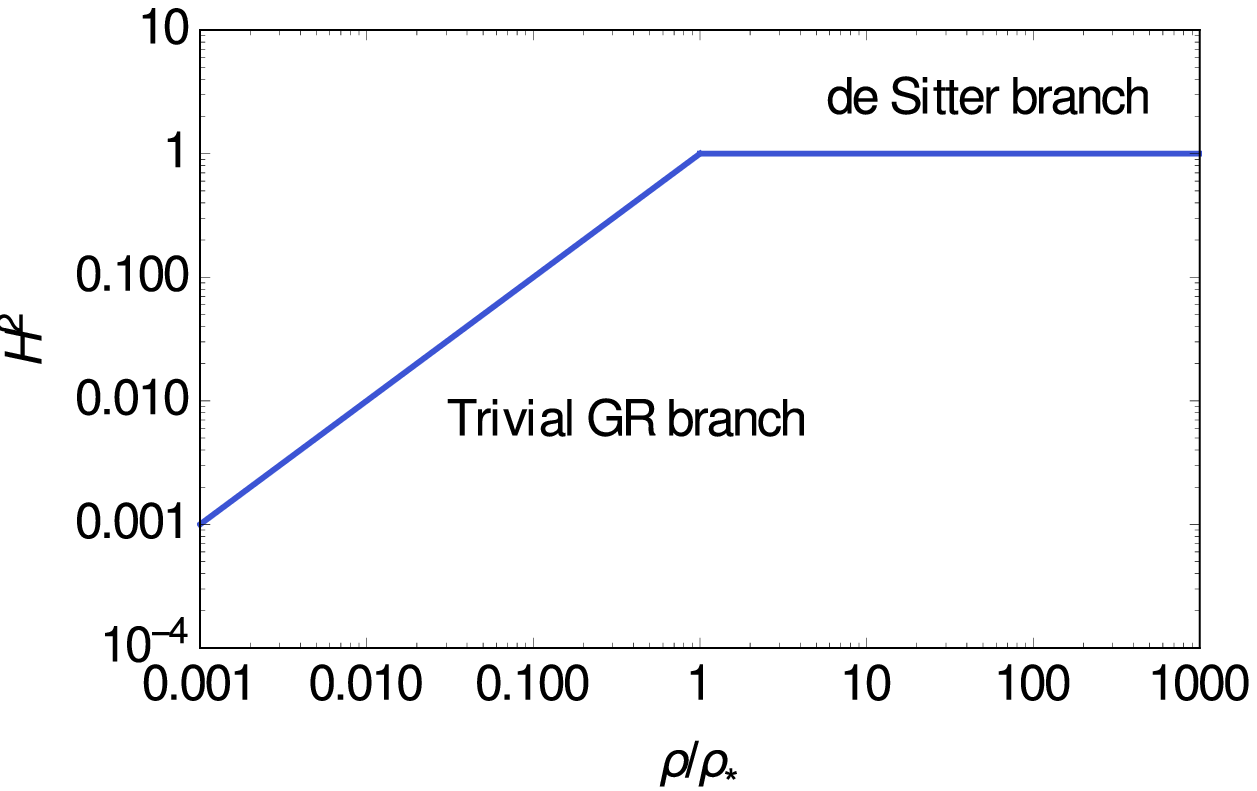}
\caption{Model $\lambda=\xi=0$. In the left panel we show the transition from the de Sitter branch with $A_0$ compensating for the matter fields to the trivial branch where $A_0$ vanishes and we recover the usual GR Friedman equation. We can see the sharp transition between both regimes discussed in the main text. In the right panel we show the behaviour of the Hubble expansion rate for this solution, where we can see the transition from the GR regime for $\rho<\rho_*$ to a de Sitter universe at densities $\rho>\rho_*$.}
\label{Fig:lambda0xi0}
\end{center}
\end{figure}

A singular case is the one with $\lambda=\xi=0$, since in that case the non-trivial branch of the vector field equation determines the Hubble function to be $H^2=M^2/(6\beta)$, which further implies that $M^2$ and $\beta$ must have the same sign. The value of $A_0$ will then be determined by the Friedmann equation
\be
A_0=\pm\sqrt{\frac{\rho-3\mpl^2 H^2}{6\beta H^2}}=\pm\mpl\sqrt{\frac{1}{2\beta}\left(\frac{\rho}{\rho_*}-1\right)}\,,
\ee
%with $\rho_*\equiv \mpl^2M^2/(2\beta)$. In pure vacuum with $\rho=0$ only the case $\beta<0$ leads to real solutions and we have $A_0^2=\frac{\mpl^2}{2\vert\beta\vert}$.For this solution we have that tensor perturbations are always stable since all the dependence on the the parameters of the action drops. The vector sector has propagation speed given by $c_v^2=1-2\beta$, so they are also stable for $\beta<0$. Finally, for the scalar sector we find
%\be
%\mathcal{K}_s=\frac{\big[12M^2-(36+k^4)\mpl^2\big]^2}{9M^2\mpl^2(k^2-4a^2M^2)}k^2\quad\quad{\rm and}\quad\quad
%\mathcal{V}_s=\frac{\big[12M^2-(36+k^4)\mpl^2\big]^2\big[(1+8\beta)k^2+4(1-8\beta)M^2a^2\big]}{27M^2\mpl^2(k^2-4a^2M^2)^2}k^2.
%\ee
%Thus, for $M^2<0$ (as required to have a real Hubble factor for $\beta<0$, $\mathcal{K}_s$ is positive and ghosts are not present. 
with $\rho_*\equiv \mpl^2M^2/(2\beta)$. We see that this solution features the existence of a lower or an upper bound for the energy density for positive and negative $\beta$ respectively. Thus, in this case the vector field can compensate for the matter density in order to maintain a de Sitter universe. However, the presence of a bounded allowed region for the density of the matter component prevents to screen an arbitrarily large or small (depending on the sign of $\beta$) density. This observation leads to the interesting possibility of a transition between branches. If $\beta>0$, then above the minimal density $\rho_*$ we have a de Sitter phase with $A_0$ compensating for the matter source. When we reach the minimal density, $A_0$ transits to the trivial branch with $A_0=0$, while recovering the usual Friedmann equation $3\mpl^2H^2=\rho$. However, the transition between branches is not smooth because although it is continuous, a discontinuity in the derivatives will appear (see Fig. \ref{Fig:lambda0xi0}).

\subsection{Case $\lambda=\beta=0$: Vector dark energy}
\begin{figure}
\begin{center}
\includegraphics[width=8cm]{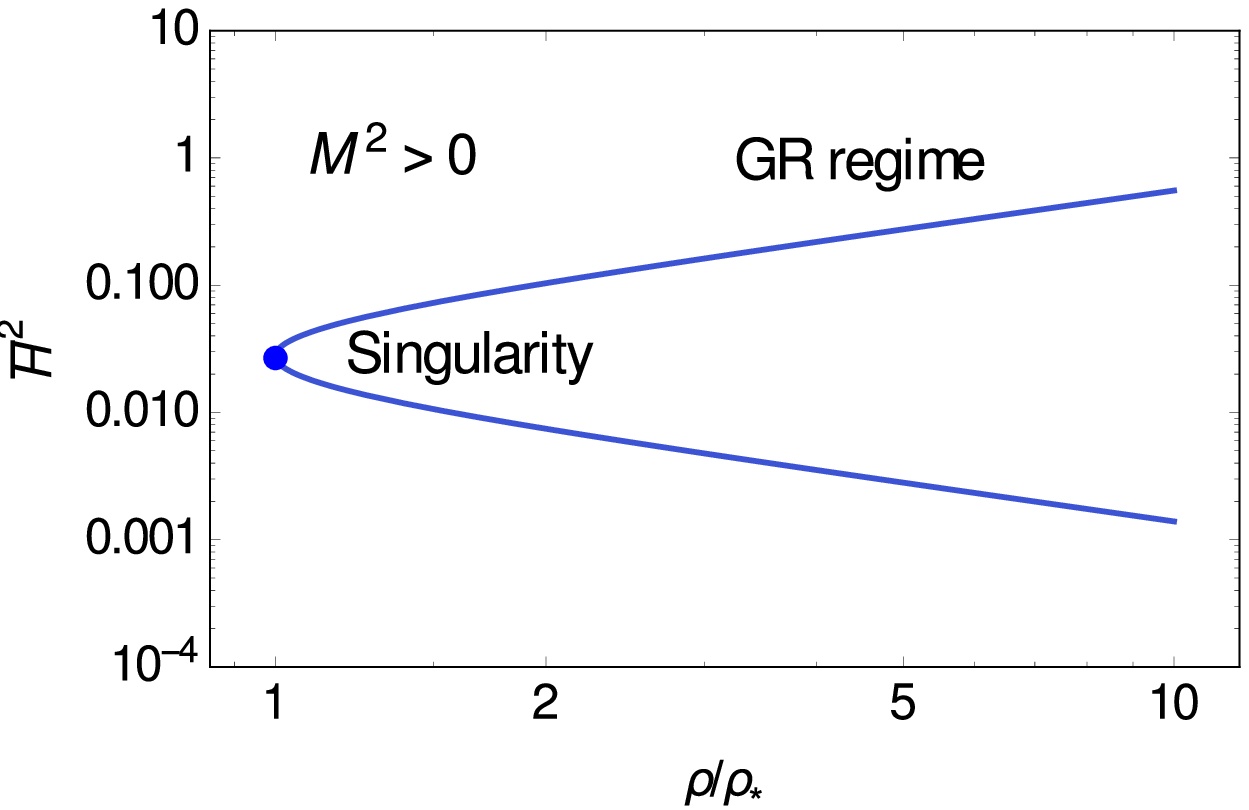}\hspace{0.75cm}
\includegraphics[width=8cm]{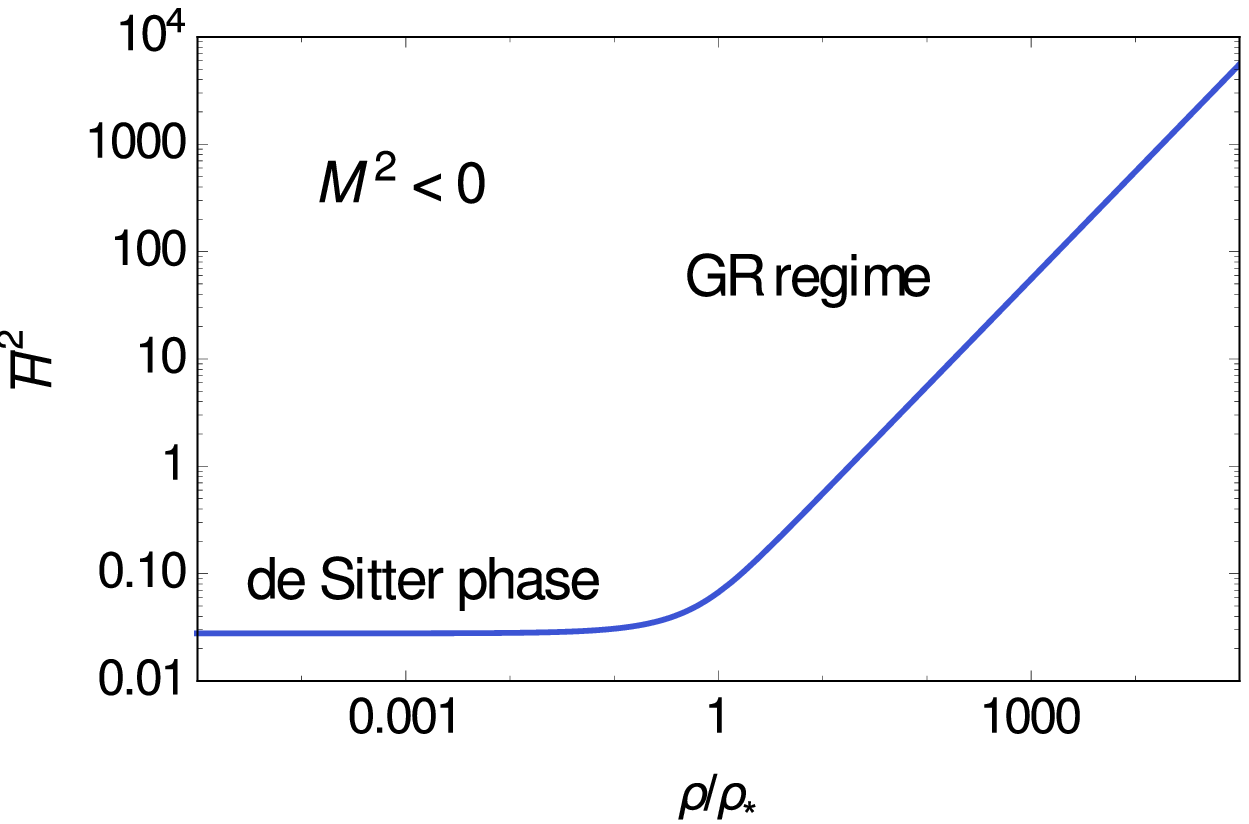}
\caption{Model: $\lambda=\beta=0$. The value of $H^2$ is normalized as $\bar{H}=\mpl^2H^2/\rho_\star$. The left panel shows the case with positive $M^2$ and the right panel shows the case with negative $M^2$. We see that for $M^2>0$ there is a minimal allowed value for the energy density for both branches, while for negative $M^2$ we have the transition from the usual GR regime at high densities to a de Sitter phase at low densities, as discussed in the main text.}
\label{Fig:lambda0beta0}
\end{center}
\end{figure}

In this case the vector field equation becomes linear so we only have one branch (besides the trivial one) with
\be
A_0=-\frac{M^2}{6\xi H}\,.
\ee
As explained above, we can further fix $\xi=1$ without loss of generality (or, equivalently, rescaling $A_0\rightarrow \xi^{1/3}A_0$ and $M^2\rightarrow \xi^{4/3}M^2$). Friedmann equation then gives the following two branches:
\be
3\mpl^2H^2_{\pm}=\frac{\rho}{2}\left(1\pm\sqrt{1-\frac{M^6\mpl^2}{6\rho^2}}\right)
\ee
If $M^2>0$, we see again the existence of a minimal density given by $\rho_{\star}=M^3\mpl/\sqrt{6}$. However, if $M^2<0$, the energy density can take any value. In the limit of small densities we then obtain
\bea
H^2_{\pm}\simeq \pm\frac16\sqrt{\frac{-M^6}{6\mpl^2}}+\frac{\rho}{6\mpl^2}\,.
\eea
On the other hand, for high densities we find
\bea
H^2_+&\simeq&\frac{\rho}{3\mpl^2}-\frac{M^6}{72\rho}\,,\nonumber\\
H^2_-&\simeq&\frac{M^6}{72\rho}\,.
\eea
We then see that the branch $H_+$ has a transition from the usual GR behaviour at high densities to a de Sitter universe at low densities. Notice that such a solution is only possible for $M^2<0$, which also guarantees that the effective cosmological constant is positive. The solutions are illustrated for both negative and positive $M^2$ in Fig. \ref{Fig:lambda0beta0}.

\subsection{Case $\lambda=0$: modified early cosmology}

\begin{figure}
\begin{center}
\includegraphics[width=8cm]{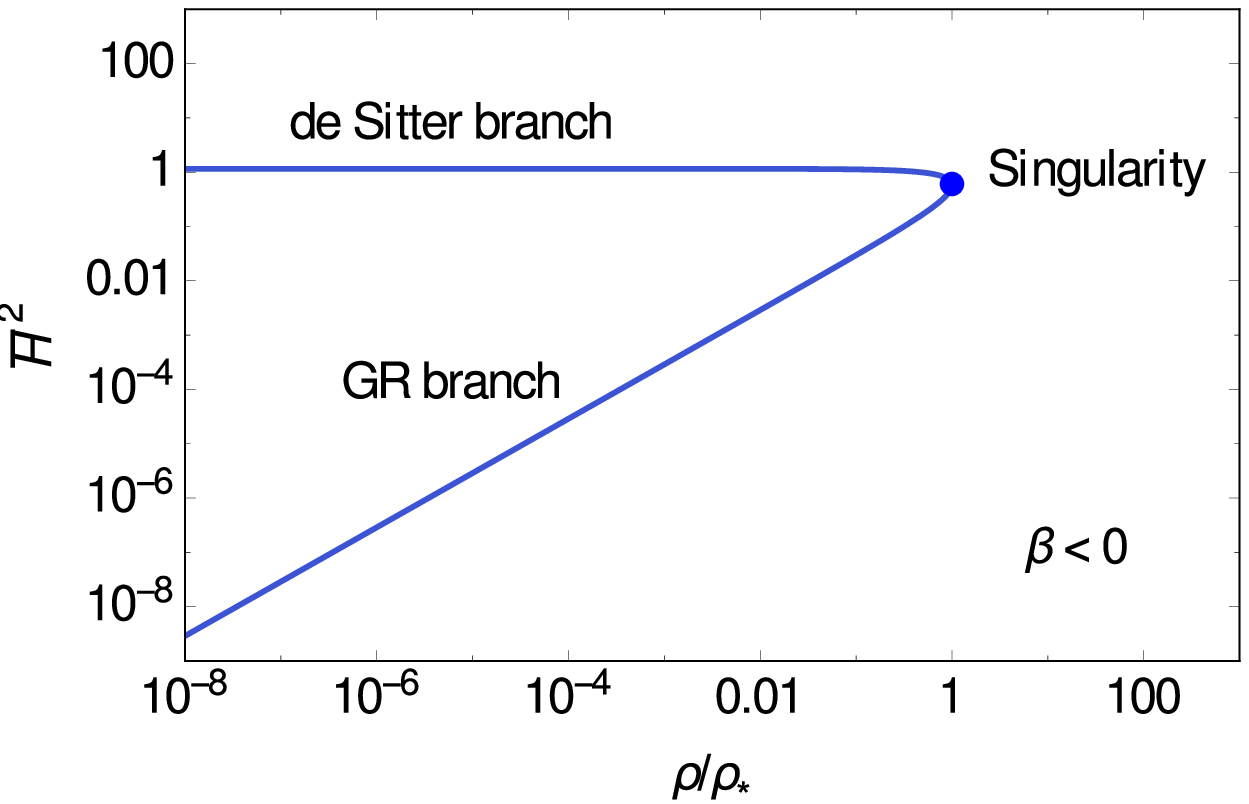}\hspace{0.75cm}
\includegraphics[width=8cm]{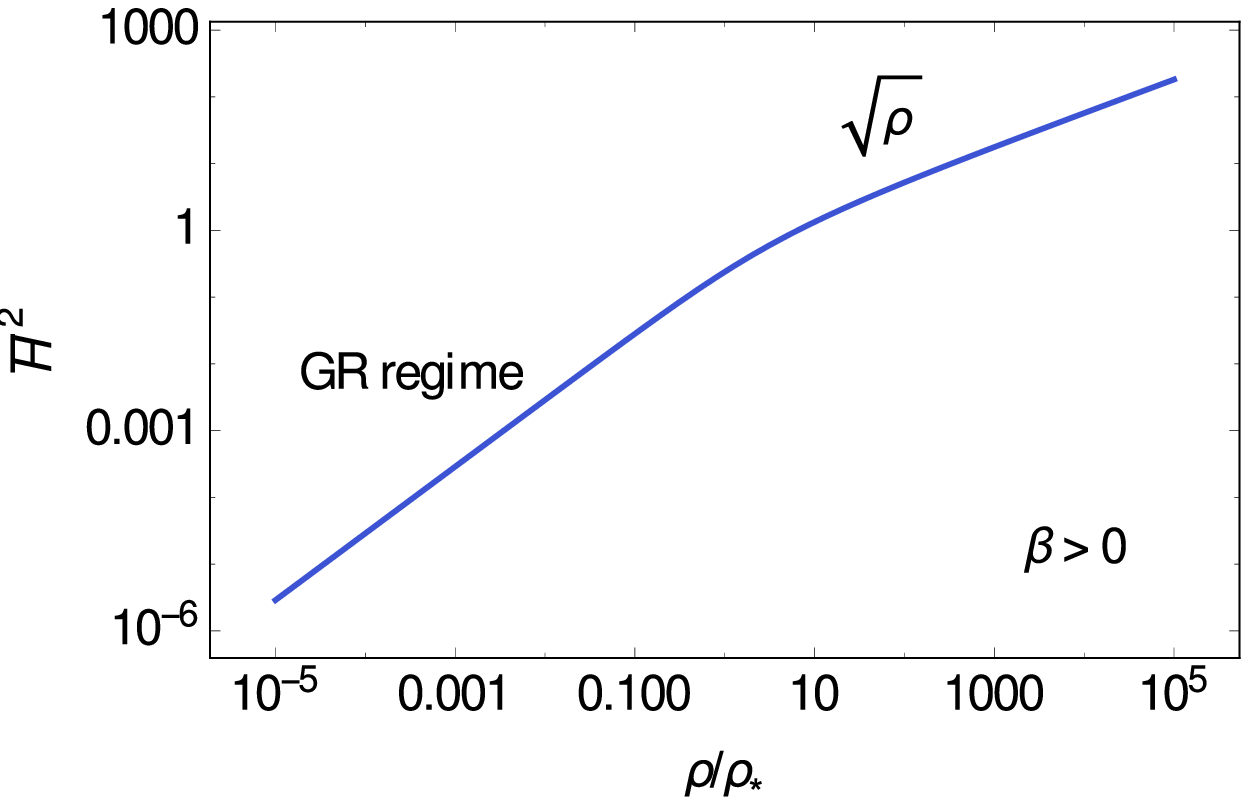}
\caption{Model: $\lambda=M^2=0$. The value of $H^2$ is normalized as $\bar{H^2}=\vert\beta^3\vert H^2/\mpl^2$. The left panel shows the case with negative $\beta$ and the right panel shows the case with positive $\beta$. We see the existence of an upper energy density for $\beta<0$. In that case both branches exist, one mimicking GR in the range of allowed energy density while in the second branch we find a de Sitter universe irrespective of the value of $\rho$. In the case with positive $\beta$ only one branch exists, reproducing GR at low energy densities, while at high energy densities we have $H^2\propto \sqrt{\rho}$. }
\label{Fig:lambda0}
\end{center}
\end{figure}

The vector field equation yields in this case
\be 
A_0=-\frac{M^2}{6\xi H}+\frac{\beta}{\xi}H\,.
\ee
Again, without loss of generality, we will set $\xi=1$. From the Friedmann equation we obtain
\be
3\mpl^2H^2+\frac{(M^2-6\beta H^2)^2(M^2+6\beta H^2)}{72H^2}=\rho\,.
\ee
The second term in the LHS going as $M^4/H^2$ of this effective Friedman equation indicates that only if $M^2=0$ we can recover the usual GR equation in the small-$H$ regime (we disregard the possibility $\beta=0$ as we would reduce to the case considered in the previous section). Thus we will set $M^2=0$. Then, we can rewrite the above expression as
\be
3\mpl^2H^2\left(1+\beta^3\frac{H^2}{\mpl^2}\right)=\rho\quad\Rightarrow\quad H^2_\pm=-\frac{\mpl^2}{2\beta^3}\left(1\pm\sqrt{1+\frac{4\beta^3\rho}{3\mpl^4}} \right)\,.
\ee
Thus, if $\beta<0$, there is a maximum energy density given by $\rho_{\star}=3\mpl^4/(4\vert\beta\vert^3)$.  Also notice that in the case with $\beta<0$, we automatically have that $H^2$ is positive. Interestingly, we see that when such a maximum density is reached, the Hubble rate remains finite in both branches: $H^2_\pm\rightarrow -\mpl^2/\beta^3$, although the derivative diverges $\vert\dot{H}_\pm\vert\rightarrow \infty$, in accordance with our general discussion on the singularities. At low energy densities $\rho\ll \rho_{\star}$, we find the usual GR result in the branch $H_-$, while the branch $H_+$ gives a de Sitter universe  with $H_+^2=-\mpl^2/\beta^3$ irrespective of the value of $\rho$. In fact, these behaviours are effective up to energy densities very close to $\rho_{\star}$ (see Fig. \ref{Fig:lambda0}). For the de Sitter branch at low densities, we can use our results on the perturbations around the de Sitter fixed points in vacuum. 

When $\beta>0$, only the branch $H_-$ exists, since $H_+^2$ is negative in that case. We find again two regimes determined by $\rho_\star$. At low densities $\rho\ll\rho_\star$, we recover the GR regime with the usual Friedman equation, i.e., $3\mpl^2H_-^2\simeq\rho$. However, in the high energy density regime we have 
\be
H^2_-(\rho\gg\rho_*)\simeq\sqrt{\frac{\rho}{3\beta^3}}\,. 
\ee
This behaviour leads to an evolution of the scale factor $a\propto t^{\frac{4}{3(1+w)}}$ so that we have accelerated expansion for $w<1/3$. In particular, for dust we have that the first slow roll parameter is $\epsilon =-\dot{H}/H^2=3/4$. In principle, this would be in tension with CMB observations, but we should remember that since we are not in the GR regime, scalar perturbations do not necessarily acquire a spectral index determined by $\epsilon$. Another interesting feature of this regime is that the effective Friedmann equation does not contain any dimensionfull constant since the dependence on the Planck mass of the full equation drops in this limit.

\subsection{Case $\beta=\xi=0$: Vector cosmological constant} 

Let us finally look at the perhaps the simplest non-trivial corner of the 4-dimensional parameter space.
The nontrivial solution for the vector is then just the constant $A_0^2=M^2/4\la$. Thereby model can reproduce the $\Lambda$CDM background, with the $\Lambda$ given by the constant vector field as
\be
3\mpl^2H^2 = \rho + \Lambda_{\la M}\,, \quad \text{where} \quad \Lambda_{\la M} \equiv -\frac{M^4}{16\la}\,.
\ee
Obtaining consistently an effective positive cosmological constant requires $M^2<0$ and $\lambda<0$.
The vector field fluctuates, and thus model could be distinguished, at least in principle, from $\Lambda$CDM by studying perturbations. In the asymptotic future de Sitter limit, both tensor and vector perturbations acquire the standard form for their quadratic lagrangians with kinetic term $\mathcal{K}_{v,t}=1$ propagation speed $c_{v,t}^2=1$ as consequence of having $\beta=0$. For the scalar sector, the kinetic and potential coefficients defined in (\ref{kmatrix},\ref{vmatrix}) are given by 
\be
{\mathcal K}_{s} =  \frac{8(-3M^2+37\mpl^2)^2}{-9M^2\mpl^2(1-2M^2a^2/k^2)}\,, \quad {\mathcal V}_{s} = 0\,.
\ee
Since the kinetic coefficient is positive definite when $ \Lambda_{\la M}>0$ (implying $M^2<0$), ghost-like instabilities are avoided. The propagation speed vanishes identically so no Laplacian instabilities will arise neither. However, this is a very singular case and means that the scalar perturbations will not actually propagate at this order and it would be necessary to include some higher order terms. 

%\subsection{Case $\xi=0$}
%The $\xi$-term in the action is the responsible for the appearance of the linear terms in $H$ in the vector field equations and its corresponding energy density. Therefore, if $\xi=0$, the equations only contain $H^2$ and $A_0^2$. The solution for $A_0$ is simply
%\be
%A_0^2=\frac{M^2-6\beta H^2}{4\lambda}.
%\ee
%We see that the existence of real solutions for any $H$ is guaranteed if $M^2$ and $\lambda$ are positive and $\beta$ is negative. If these conditions are not satisfied, there will be a bound on the possible values for $H$.
%
%After solving for $A_0$ and plugging the solutions into the Friedmann equation, we obtain
%\be
%3\mpl^2\left(1+\frac{\beta M^2}{4\lambda \mpl^2}\right)H^2-\frac{27\beta^2}{4\lambda}H^4=\rho-\frac{M^4}{16\lambda}
%\ee
%which can be inverted to express $H^2$ as a function of $\rho$. The existence of real solutions will be determined by the discriminant of the equation, which can be expressed as
%\be
%\Delta\equiv \left(1+\frac{\beta M^2}{4\lambda \mpl^2}\right)
%\ee

\section{Discussion and conclusions}
\label{conclusions}

In this paper we have studied cosmological solutions for a class of vector-tensor theories that arise within the framework of the geometries introduced in \cite{Jimenez:2014rna,Jimenez:2015fva}. Those geometries are defined by the most general connection linearly determined by a vector field (and without derivatives). We have then considered the most general action quadratic in the Riemann curvature tensor for this connection. Remarkably, we found that the absence of ghostly degrees of freedom associated to more than 3 polarizations for the vector field restricts the connection to be of the extended Weyl type. This extended version of Weyl geometry maintains the conformal invariance of the non-metric compatibility while, in addition to non-metricity, allowing the presence of the vector trace torsion. The resulting vector-tensor theory contains the usual Maxwell kinetic term with a mass and a quartic potential. Moreover, we also found a cubic derivative self-interaction for the vector field which gives the cubic Galileon in the decoupling limit and a non-minimal coupling to the Einstein tensor. Although in this work we have focused on the quadratic theory, we can argue that higher order derivative self-interactions with Horndeski vector-tensor and Galileon-like terms for the St\"uckelberg field in the decoupling limit can also be constructed by considering cubic and higher order curvature terms in the action.

For the vector-tensor theory arising from quadratic curvature terms, we have shown the existence of isotropic de Sitter solutions. These solutions are supported by the temporal component of the vector field, which is non-dynamical and plays the role of an auxiliary field. For the general case, we have obtained the regions in the parameter space of the theory where such solutions exist and performed a classification according to the number of de Sitter branches. We have then studied the stability of the inhomogeneous perturbations around the de Sitter critical points by computing the quadratic action for the scalar, vector and tensor perturbations. This analysis explicitly showed that the theory only propagates the 2 polarizations of the graviton plus the 3 polarizations of the massive vector field. For vector perturbations the only effect is a modification of the propagation speed so that they are never ghostly and only a potential Laplacian instability can be present. Furthermore, the effects in both the vector and the tensor sectors vanish if the coupling to the Einstein tensor is turned off. For the scalar sector there is a modification of the kinetic term and the propagation speed that depends on all the terms in the action, so the scalar mode can potentially contain both ghost and/or Laplacian instabilities around the de Sitter solutions.

After devoting a careful analysis to the de Sitter solutions we have turned to the general cosmological evolution in the presence of a matter sector. We have shown that, typically, we might expect to have upper and lower bounds for both the Hubble expansion rate and the energy density. We have also considered the possibility of having bouncing solutions and found that this is only potentially possible in a stable way if the vector Galileon interaction is present. In any case, we have shown that the effective Newton's constant in the background Friedmann equation must be negative at the bounce, although this by itself does not represent an instability. However, we ahve shown that bouncing solutions at high energies generically cannot match continuously a GR regime at low densities. Moreover, we have also studied the presence of turnovers with a transition from an expanding to a contracting universe. Another type of turnovers that we have analyzed is those happening at a finite $H$ so that they cannot correspond to the discussed transtions. We have shown that these additional turnovers correspond to a cosmological evolution where $\dot{H}$ diverges. We have also analyzed the existence of self-tuning solutions such that we can have Minkowski solutions in the presence of an arbitrary cosmological constant. As in the Horndeski class of theories, this is only possible for non-flat FLRW universes for our vector-tensor theory. Moreover, such solutions can be realized if only the coupling to the Einstein tensor is present, with all the other terms vanishing.

Finally, we have considered some specific examples to illustrate the general results discussed in the first part of the paper. We have shown that the class of vector-tensor theories studied in this work and which naturally arise in the framework of geometries with a linear vector distortion can give rise to a rich an interesting phenomenology for cosmology. The effects of the vector field can be important both in the early universe or at late times depending on the values of parameters. Therefore, they can be used to build dark energy/dark matter models or inflationary scenarios. Moreover, the natural presence of upper bounds for the energy density and/or the curvature might help evading singularities without resorting to quantum effects. In this work we have focused on purely isotropic universes, but the spatial components of the vector field and the class of anisotropic solutions that may be obtained could also deserve attention. 
%In particular, what new types of cosmological solutions can be obtained. W leave this for future work.

\vspace{0.5cm}
{\bf Acknowledgments}: We thank Federico Piazza for useful discussions. JBJ  acknowledges  the  financial  support  of
A*MIDEX project (n ANR-11-IDEX-0001-02) funded by
the  "Investissements d'Avenir" French Government program, managed by the French National Research Agency
(ANR),  MINECO  (Spain) projects
FIS2014-52837-P and Consolider-Ingenio MULTIDARK
CSD2009-00064. L.H. acknowledges financial support from Dr. Max R\"ossler, the Walter Haefner Foundation and the ETH Zurich Foundation.

\appendix

\section{The $\gamma_i$'s}

In this appendix we will give some details on the derivation of our vector-tensor theory and how the generalized Weyl geometries stand out as the ones leading to theories free from Ostrogradski instabilities. The relevant coefficients for that are $\gamma_1$, $\gamma_2$ and $\gamma_3$, whose explicit expressions are
\begin{align}
\hspace{-2cm}\gamma_1=\frac{b_3+b_2-2b_1}{4}(D-1)\Big[&(b_2+b_3)\Big(5-8c_3-8 c_4-4c_5-4c_6-d_2-2d_3-3D+ 
 4(c_3+c_4+c_5+c_6)D\Big)\nonumber\\
 &+2 b_1\Big(3-4c_5-4c_6+d_2+2d_3+4c_3(-2+D)+4c_4(D-2)-D\Big)\Big]\,,
\end{align}
\begin{align}
\gamma_2=\frac{b_3+b_2-2b_1}{4}\Big[&(b_2+b_3)\Big(14 - 8 c_5 - 8 c_6 - 4 (c_3+c_4) (D-2)^2 - (14 - 8 c_5 - 8 c_6 - d_2 - 2 d_3) D + 
 4 (1 - c_5 - c_6) D^2 \Big)\nonumber\\
 &-2b_1\Big(   4 (c3+c_4) (D-2)^2  + (-8 c_5 - 8 c_6 + d_2 + 2 d_3) D + 
 2 (-1 + 4 c_5 + 4 c_6 + D)  \Big)\Big]\,,
\end{align}
\begin{align}
\hspace{-8.9cm}\gamma_3=(b_3+b_2-2b_1) (1 - 2 c_3 - 2 c_4 - c_5 - c_6) (D-2)\,.
\end{align}
We can immediately see that these coefficients identically vanish for the generalized Weyl geometries with $b_3=2b_1-b_2$. One would expect that, given the numerous independent coefficients available, additional solutions to the equations $\gamma_i=0$ exist for any $D$. In the following we will show that this is not the case.  Let us assume that $b_3\neq 2b_1-b_2$ and solve $\gamma_3=0$ for $c_6$ to obtain $c_6=1 - 2 c_3 - 2 c_4 - c_5$. If we plug this solution into $\gamma_1$ and $\gamma_2$ we obtain:
\begin{eqnarray}
\gamma_1&=&\frac{ (b_3+b_2-2b_1)^2}{4} (D-1) \Big[1 - d_2 - 
   2 d_3 + (1 - 4 c_3 - 4 c_4) D\Big]\,,\\
\gamma_2&=&\frac{ (b_3+b_2-2b_1)^2}{8} \Big[6 + D \Big(-6 + d_2 + 2 d_3 + 4 (c_3 + c_4) D\Big)\Big]\,.
\end{eqnarray}
Now we can solve $\gamma_1=0$ for any of the remaining parameters, say $d_2$, and insert it in the expression for $\gamma_2$ which finally reduces to
\be
\gamma_2=\frac{ (b_3+b_2-2b_1)^2}{4D}(D-3)(D-2)(D-1)\,.
\ee
Thus, we see that for $D\geq 4$ there are no additional solutions besides the generalised Weyl geometry with $b_3=2b_1-b_2$. We can also see that in lower dimensions additional solutions exist.

\bibliography{modgrav2}

\end{document}